\documentclass[12pt,letterpaper]{article}         
\usepackage{osajnl}
\usepackage{overcite}
\usepackage{epsfig, array, amsmath, delarray}

\begin{document}
\baselineskip=.175in


\title{Analytical Modeling of the White Light Fringe}

\author{Slava G. Turyshev}

\affiliation{Jet Propulsion Laboratory, California Institute
of  Technology, Pasadena, CA 91109 }

\vspace*{1.1cm}

\begin{abstract}
We developed analytical technique for extracting the phase, visibility and
amplitude information as needed for interferometric astrometry with the Space
Interferometry Mission (SIM).  Our model accounts for a number of physical
and instrumental effects, and is valid for a general case of bandpass filter. We were
able to obtain general  solution for polychromatic phasors and address properties of
unbiased fringe estimators in the presence of  noise.  For
demonstration purposes we studied the case of rectangular bandpass filter with two
different methods of optical path difference (OPD) modulation -- stepping  and
ramping  OPD modulations. A number of areas of further studies relevant
to instrument design and simulations  are outlined and discussed.  

\end{abstract}
\ocis{120.2440, 120.2650, 120.3180, 120.5050, 120.5060}


\maketitle 

\section*{Introduction}
\label{sec:intro}

SIM is designed as a space-based 10-m baseline Michelson optical interferometer
operating in the visible waveband (see Ref. \citeonline{sim} for more details). This
mission will open up many areas of astrophysics, via astrometry with unprecedented
accuracy.  Thus, over a narrow field of view SIM is expected to achieve mission
accuracy of 1 $\mu$as.  In this mode  SIM will search for planetary companions to
nearby stars  by detecting the astrometric ``wobble'' relative to a nearby ($\le
1^\circ$) reference star.  In its wide-angle mode, SIM will be capable to provide a 4
$\mu$as~ precision absolute position measurements of stars, with parallaxes to
comparable accuracy, at the end of a 5-year mission.  The expected proper motion
accuracy is around 3 $\mu$as/yr, corresponding to a transverse velocity of 10~m/s
at a distance of 1~kpc.
 
\begin{figure}[ht]
\centering\epsfig{file=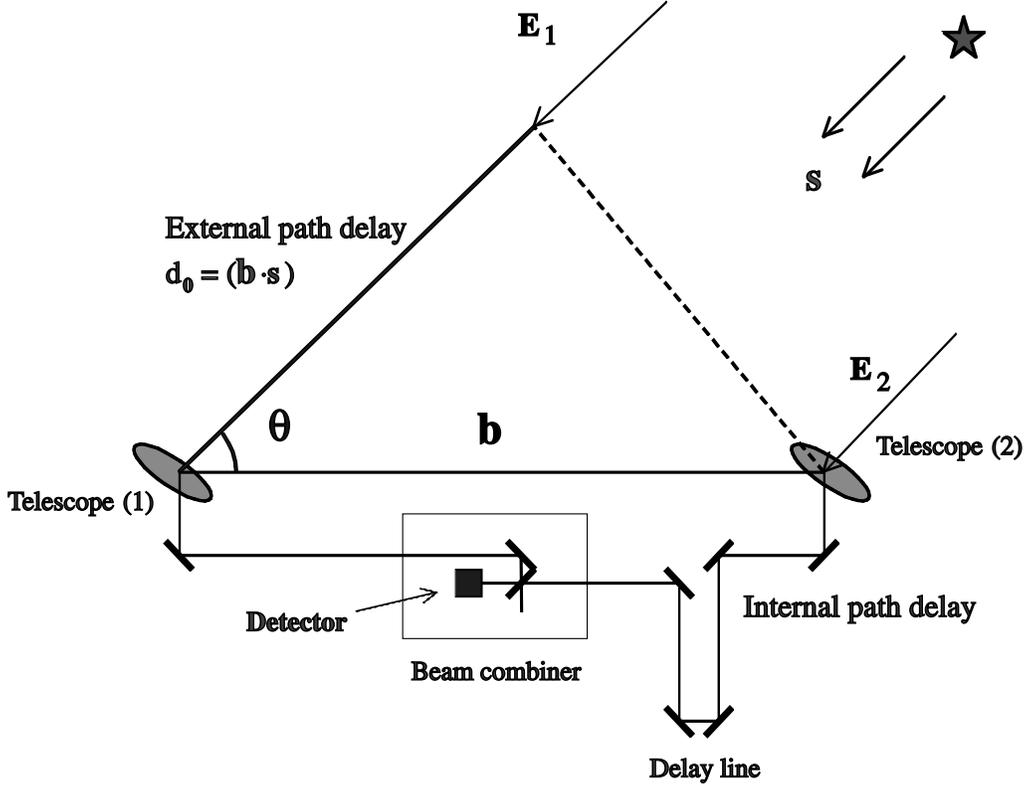,width=13.5cm}
\caption{Basic geometry of light propagation in the stellar interferometry.}
\label{fig:astrom}
\end{figure}

The SIM instrument does not directly measure the angular separation between stars,
but the projection of each star direction vector onto the interferometer baseline
by measuring the pathlength delay of starlight as it passes through the two arms of
the interferometer. The delay measurement is made by a combination of internal
metrology measurements to determine the distance the starlight travels through each
arm,  external metrology measurements that determine the length and local
orientation of the baseline, and a measurement of the central white light fringe to
determine the point of equal optical pathlength (see Figure \ref{fig:astrom}).  

The primary motivation for the work was the idea to use the averaged and 
bias-corrected complex phasors to estimate the external optical pathlength difference
for the incoming polychromatic light. This approach is applicable to the science
interferometer phase measurement, for which it is not necessary to remove the bias
from each estimate, just as long as it is removed from the averaged estimate.  This
method was applied in Ref. \onlinecite{Quirrenbach} to obtain unbiased fringe
visibility estimates. Under certain conditions (dc intensity values and visibilities
approximately constant during integration time) this method allows to use the average
values of the phasors for estimating the average phase
$\bar{\phi}$ (see discussion in Refs. \onlinecite{Turyshev2000,markscott}). A
definition for the complex visibility phasors is stemming from the form of a complex
visibility function, $\tilde V =ve^{j\phi}$, where $v$ is the visibility  and $\phi$
is its phase. Decomposing this expression onto real and imaginary parts as $\tilde V
=X+jY$, one obtains the complex visibility phasors,  $X=v\cos\phi$ and $Y=v\sin\phi$.
SIM will be able to effectively determine both visibility and phase of the fringe, but
for the astrometric purposes the phase must be determined to a much higher accuracy.
(The most stringent SIM requirement in this regard is the average phase error over 30
s integration time corresponds to a path-length error of approximately 30 pm.)
 Phase determination in the presence of noise is a non-linear estimation process. Even
in the monochromatic case it  requires careful approach to averaging and correcting
for biases in the data. Most of the algorithms for phase-shifting interferometry are
designed for monochromatic light, and there is typically a match between the stroke of
the modulating element and the wavelength of the light (see, for example, Refs.
\onlinecite{creath} for a discussion of various algorithms). However, because SIM uses
a dispersed fringe technique, this match cannot be maintained over multiple channels.
A number of modifications to the existing four-bucket algorithms were designed to
address specific problems relating to Palomar Testbed Interferometer
(PTI)\cite{colavita1} and the Keck Interferometer.  Also, there is nothing inherent
that requires the match between wavelength and stroke, nevertheless most analysis has
been based on this assumption. It was shown in Ref. \citeonline{markscott} that at low
light levels, most of these algorithms become biased when there is a mismatch. The
magnitude of this bias is quite significant with respect to the instrument
requirements, thus making these algorithms unacceptable for SIM. 

This paper discusses analytic model developed for the white light fringe data
extraction.  Our goal here is to establish functional dependency of the white light
fringe parameters on the properties of incoming light as well as the instrumental
input parameters. We will show that this approach is applicable for establishing the
unbiased estimators for the case with low light levels. This method enables on to
analytically analyze the noise propagation properties. The importance of this
feature comes from the fact that integration time on stellar targets accounts for
a significant portion of the mission time and can be reduced by the use of processing
techniques with reduced error variance (see
Refs.\citeonline{shao}-\citeonline{colavita2} for more details). The analytical from
of the fringe parameters may be helpful in studying different properties of the
instruments, especially their contribution to the accuracy of  astrometric delay
measured by SIM science interferometer.\cite{Boden,Swartz}  

The problem of
interference of electromagnetic radiation is well studied and  extensive number of
publications on this subject (specifically related to stellar interferometry)  are
available (see Refs. \citeonline{colavita2},\citeonline{goodman}-\citeonline{mct} and
references therein).  However, because of complexity of this problem in a general case
of polychromatic light, most of the current research is done numerically. While
numerical studies have proven to be extremely valuable in analyzing the interference
patterns and are very useful in addressing various instrumental effects, the
analytical methods provide the much needed critical understanding of the white light
interference phenomena. It will be demonstrated below that analytic solution may be
used as a  tool  to study the complex interferometric phenomena on a principally
different qualitative level.\cite{Turyshev2000} 

In this paper we derive analytic model that may be used 
to describe  photo-electron detection process. We analytically describe the
physical and instrumental processes that are important in estimating the fringe
parameters (i.e. intensity of incoming radiation, its visibility and the phase of the
fringe). Effects that are not included in the  model are due to polarization of both
incoming light and the instrumental throughput, effect of the wavefront-tilt,  low
frequency vibrations, drifts, jitter, etc. Consideration the size of this paper,
we plan to address these issues elsewhere.  

The paper is organized as follows:  In Section \ref{sec:mono} we review the
description of interferometric pattern in the case of  monochromatic radiation. In
Section \ref{sec:model} we develop a model for the interferometric pattern registered
by a CCD detector in the polychromatic case. Our model accounts for the effects of the  
instrumental throughput, beam splitter and quantum efficiency of the CCD.
We also discuss the spectral channels with narrow bands designed to filter the 
polychromatic light. In  Section \ref{sec:integr} we  introduce parameterization for
the  polychromatic fringe pattern and define the quantities that are forming the
astrometric signal on the CCD.   Specifically, we   derive solution for the white
light fringe equation in the general case.  In Section \ref{sec:pol_g} we present
general analytic solution for complex visibility phasors and will discuss a noise
suppression approach.  In Section \ref{sec:filter} we develop technique for studying
the case of a rectangular bandpass filter. We also    obtain  functional  dependency
of our solution  in the two cases of OPD modulation, namely the stepping and ramping
modulations.  In Section \ref{sec:summary} we  present conclusions and recommendations
for future studies of accurate fringe reconstruction.  In order to make  access of the
basic results  of this paper easier, we will present some important calculations in
the Appendices. Thus, in Appendix \ref{sec:app_phase} we discuss two possible
definitions for the fringe phase and justify the choice we made in the paper. In
Appendix \ref{sec:appa} we develop approximation for the complex fringe envelope
function. In Appendix \ref{sec:ow_rec} we present a general solution for the
instrumental contribution affecting the fringe parameters in the case of the
rectangular bandpass  filter and stepping OPD modulation. 


\section{Monochromatic Fringe Pattern}
\label{sec:mono}
 
The problem of interference of a monochromatic radiation presently is well studied
(see Ref.\onlinecite{goodman} and references therein). Here we would like to review
information that will be necessary for discussion of the fringe phase extraction
process.

We assume that  a plane electromagnetic wave, $\vec{E}$, which is coming from
infinity simultaneously on  the two arms of an interferometer, has the following form:
\begin{equation}
\vec{E}=\vec{E}_0e^{j\,\big(\omega t - {\vec k}\cdot{\vec x}\big)},
\label{eq:em_mono}
\end{equation}
\noindent with $\vec{E}_0$ is a constant vector. We will be using a nomenclature
where a wavenumber $k$  relates  to the wavelength as follows
$k=\frac{2\pi}{\lambda}$. After passing through the interferometer, the two beams
$\vec{E}_1$ and $\vec{E}_2$, are combined at the detector to produce   interferometric
pattern (see Figure~\ref{fig:astrom}). The corresponding fringe pattern is due to the
coherent addition of the two light-beams,
$\vec{E}=\vec{E}_1+\vec{E}_2$, and it may be expressed as follows:
\begin{equation}
 {\cal I}= \langle \vec{E}\cdot\vec{E}^* \rangle_{\sf time} = 
{\vec{E}^2}_{01}+\vec{E}^2_{02} +2\langle\vec{E}_1\vec{E}_2\rangle,
\end{equation}
where $^*$ denotes a complex conjugate quantity and $\langle..\rangle_{\sf time}$
denotes a time-averaged quantity. The first two terms on the right-hand side of this
equation   are   constant intensities of light in the two beams,
${\cal I}_1={\vec{E}^2}_{01}$ and 
${\cal I}_2=\vec{E}^2_{02}$. The third one is the interferometric pattern which, in
ideal situation,   depends only on the   optical path difference between the
two beams and  the   wavelength of the radiation, namely   $\phi_{0}=\phi_1-\phi_2=
\frac{2\pi}{\lambda}d_0\equiv kd_0$. The resulting intensity of radiation on
is given
\begin{equation}
{\cal I}(k)=   {\cal I}_0\big(1+V\cos kd_0 \big),
\label{eq:fringe}
\end{equation}
\noindent where we denote the constant part in the intensity pattern as
${\cal I}_0={\cal I}_1+{\cal I}_2$. Also, the quantity $V$ is the visibility of 
the incoming light, which, in the case of the beams intensities mismatch ${\cal
I}_1\not={\cal I}_2$, is  given by\cite{goodman,Lawson00}
 \begin{equation}
V=\frac{2 \sqrt{{\cal I}_1{\cal I}_2}}{{\cal I}_1+{\cal I}_2}{\cal V}=
\frac{2|\vec{E}_{01}||\vec{E}_{02}|}{\vec{E}^2_{01}+\vec{E}^2_{02}}{\cal V}.
\label{eq:vis}
\end{equation}
\noindent The factor ${\cal V}$ ($0\leq{\cal V}\leq1$) in this equation is the true
source visibility  (or the fringe contrast). In the case of monochromatic radiation
from an un-resolved source, when the intensities in the two arms of the
interferometer are of equal amplitudes, or 
$|\vec{E}_{01}|=|\vec{E}_{01}|= E_{0}$, the constant intensity becomes 
${\cal I}_1={\cal I}_2=\frac{1}{2}{\cal I}_0$. Therefore,  visibility reduces simply
to one, $V=1$.

\subsection{Fringe Modulation}
\label{sec:mod}

\begin{figure}[t]
\begin{center}  
\rotatebox{90}{\hskip 50pt Fringe amplitude}
\hskip -12pt
\begin{minipage}[b]{.46\linewidth}
\centering\psfig{figure=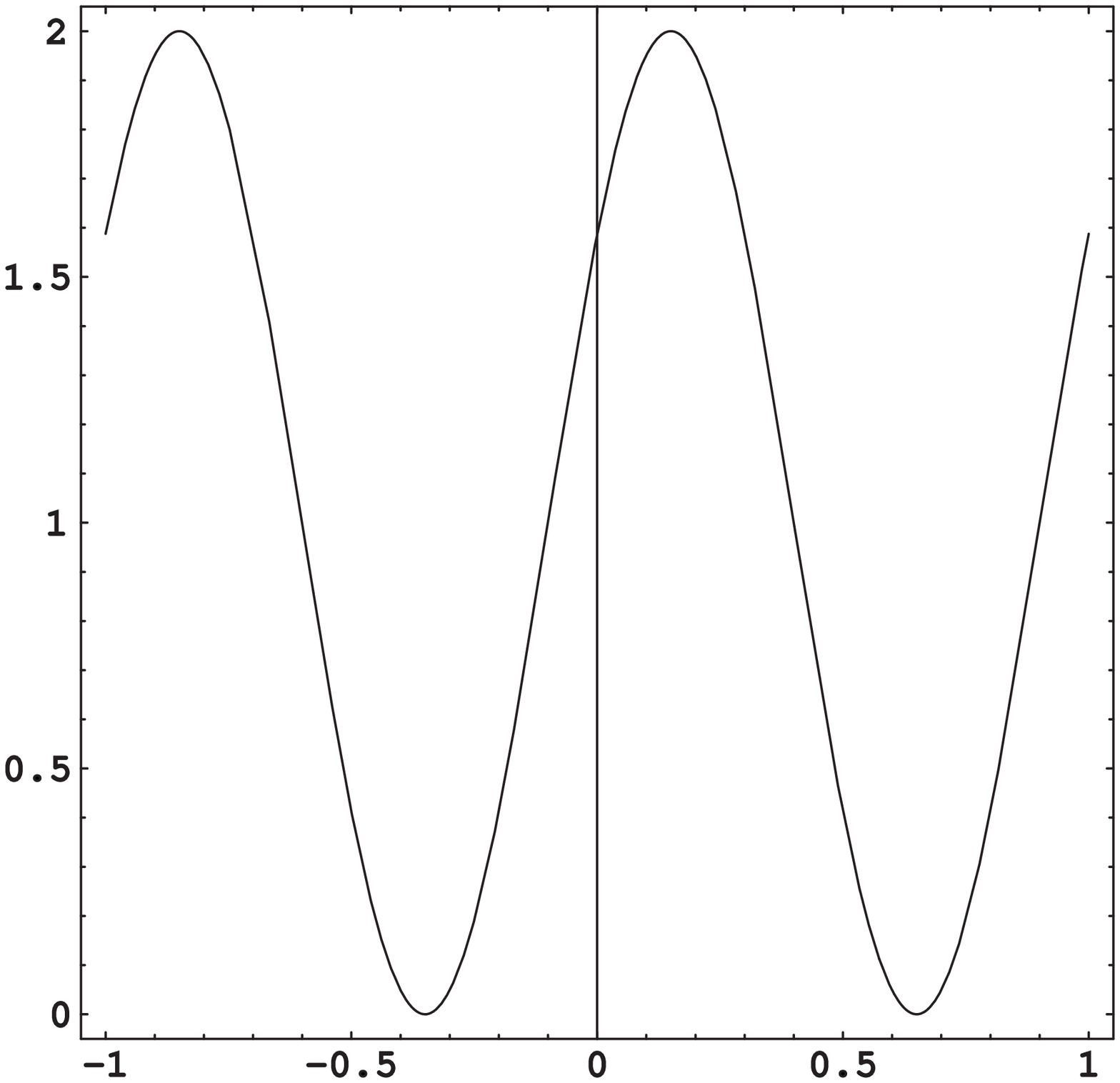,width=65mm}\\[0pt]
\rotatebox{0}{\hskip 17pt  Optical pathlength difference,~$x$ }
\end{minipage}
\caption{Monochromatic fringe. Note the visibility is set 
$V=1$ and the initial phase  offset chosen to be $\phi_0=\frac{\pi}{5}$.}
\label{fig:m_fringe}
\end{center}
\end{figure}

As seen in Figure \ref{fig:m_fringe}  monochromatic pattern is a simple one. 
In practice, the studied source may be interferometrically resolved, thus leading to a
reduced  visibility $V\not=1$. Moreover, the visibility $V$, the constant phase
difference $\phi_0$ and  the constant intensity ${\cal I}_0$ are generally  
not known. To find them one modulates the phase difference with some known
function, say $x(t)$, as below:
\begin{equation}
{\cal I}(k,t)=  {\cal I}_0\Big(1+V\sin\big( \phi_0+ k\,x(t)\big)\Big),
\label{eq:int}
\end{equation}

\noindent where effect of the beam splitter (which is true for a Michelson stellar
interferometer) brings additional $\frac{\pi}{2}$ phase shift (discussed in Section
\ref{sec:model}\ref{sec:beamsplitter}).\cite{colavita1,shao1,shao2} 

The primary goal here is to determine the phase $\phi_{0}$ by modulating the
internal optical path difference $x(t)$.  By doing so, one finds the exact value of 
internal delay $x_{\sf int}=x(t')$ that would exactly compensate the  initial offset
or $k(d_0-x_{\sf int})=0.$  Then, having determined the external delay $d_0=x_{\sf
int}$ one can determine the source position from equation $d_0=({\vec b}\cdot {\vec
s})$, where ${\vec b}$ is the interferometer baseline vector and ${\vec s}$ is the source
position on the sky. In practice, this is done in a global astrometric solutions
discussed in details in Refs.\onlinecite{Boden,Swartz} and not addressed here.

\begin{figure}[p]\vskip -20pt
\begin{center} \hskip 25pt 
\rotatebox{90}{\hskip 50pt Temporal bins}
\hskip -35pt
\begin{minipage}[b]{.9\linewidth}
\vskip 0pt
\centering\epsfig{file=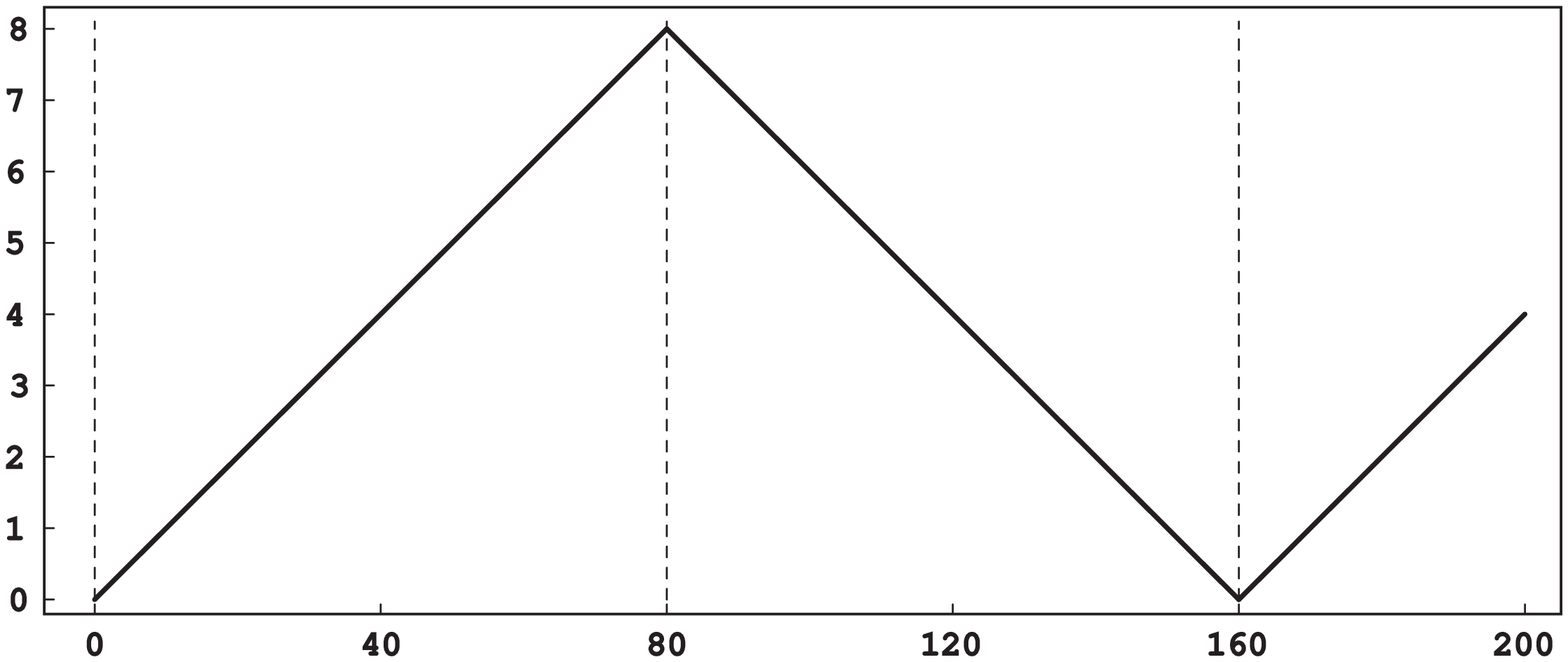,width=11cm}
\vskip -5pt \hskip -0pt
\rotatebox{0}{\hskip -10pt  Time,~[ms] }
\end{minipage}
\vskip -30pt
\hskip 15pt  
\rotatebox{90}{\hskip 50pt Fringe amplitude}
\hskip -43pt
\begin{minipage}[b]{.9\linewidth}
\vskip -25pt
\centering\epsfig{file=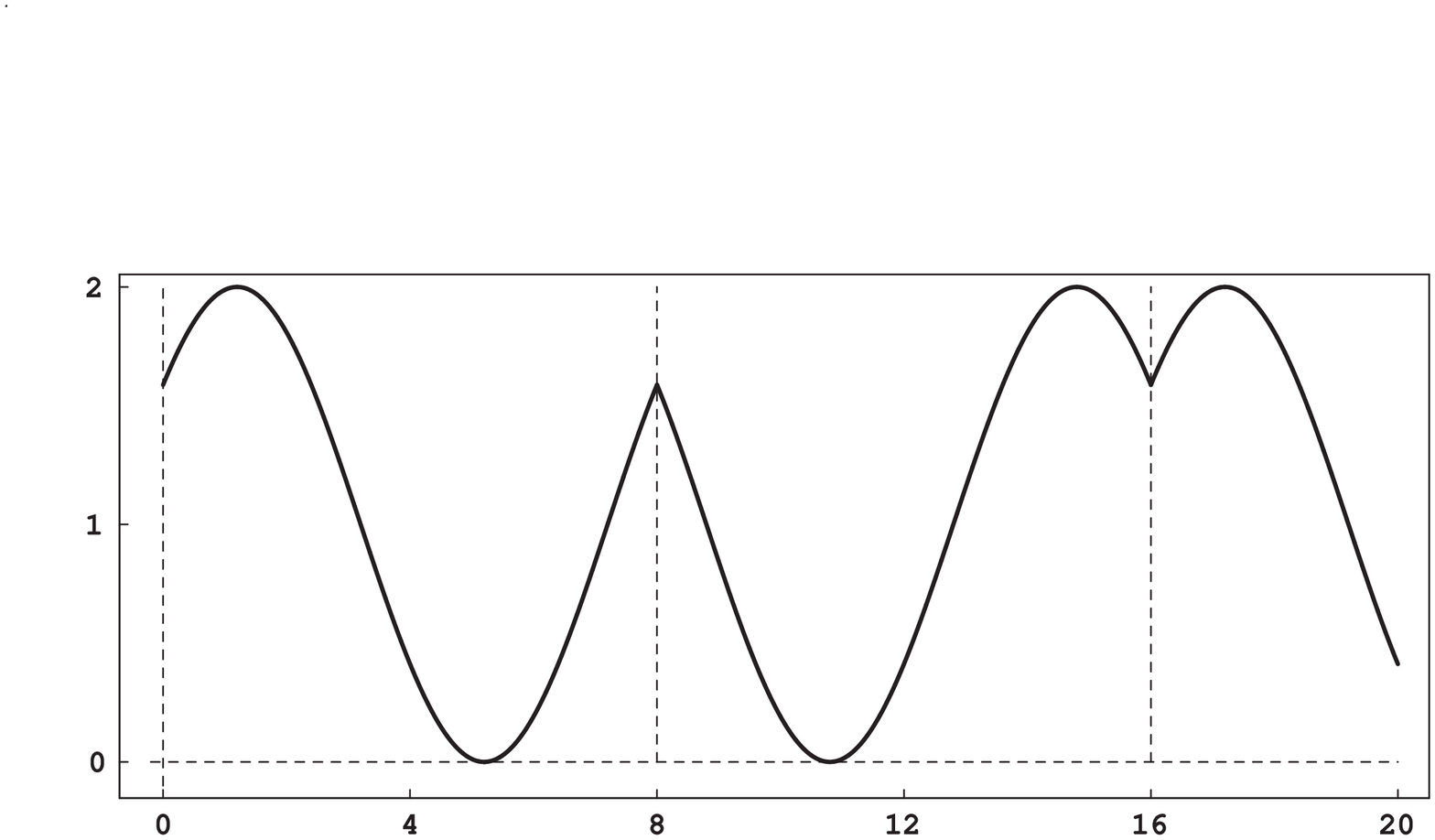,width=11.6cm}
\vskip -5pt \hskip -0pt
\rotatebox{0}{\hskip 17pt  Optical pathlength difference,~[bins] }
\end{minipage} \vskip -5pt
\caption{Upper plot shows a typical OPD modulation stroke with ramping over 8 equal
temporal bins with duration of  10 ms each. Lower plot is a monochromatic fringe as
a  function of OPD that is modulated  by ramping over a wavelength (initial phase 
offset   is  $\phi_0=\frac{\pi}{5}$ - same as in Figure \ref{fig:m_fringe}).
\label{fig:fringe_ramp}}\vskip 5pt
\end{center}
\end{figure}
\begin{figure}[p]\vskip -20pt
\begin{center} 
\hskip 25pt 
\rotatebox{90}{\hskip 50pt Temporal bins}
\hskip -35pt
\begin{minipage}[b]{.9\linewidth}
\vskip 0pt
\centering\epsfig{file=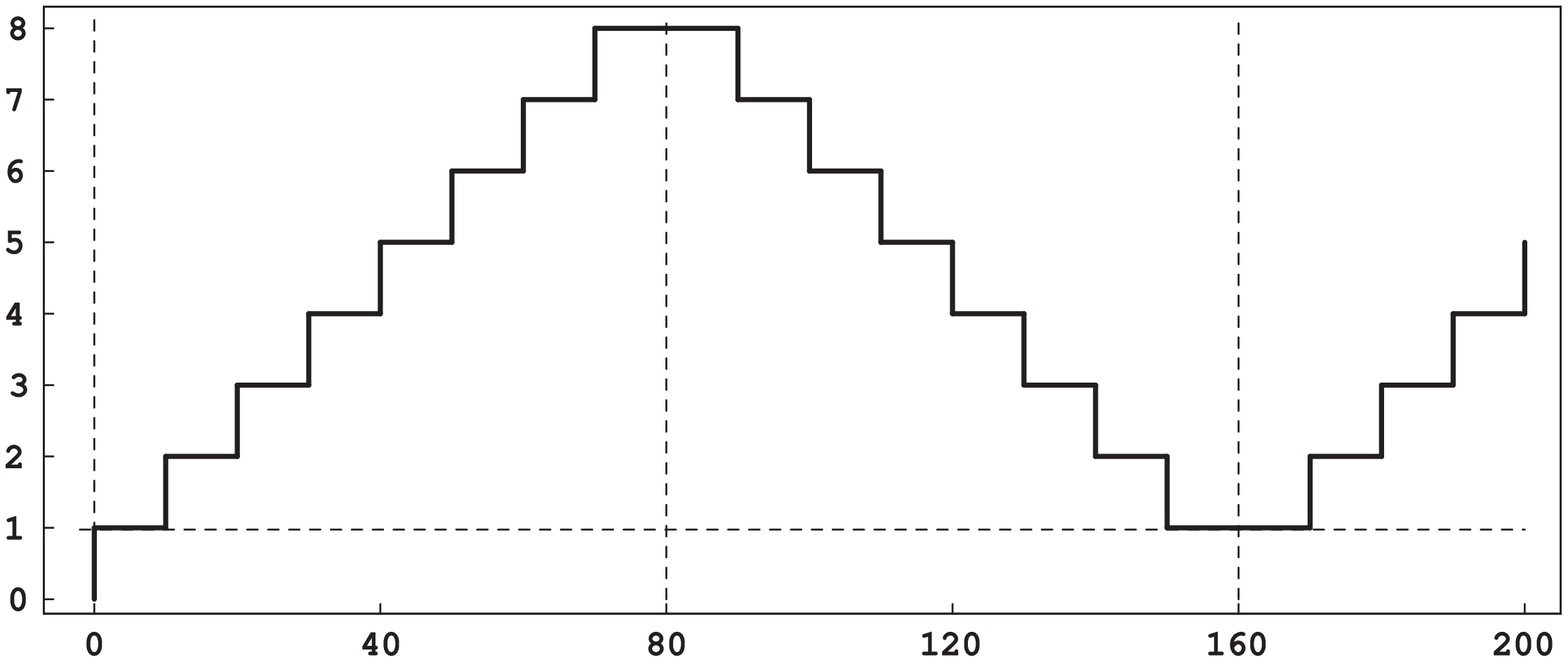,width=11cm}
\vskip -5pt \hskip -0pt
\rotatebox{0}{\hskip -10pt  Time,~[ms] }
\end{minipage}
\vskip 20pt
\hskip 25pt  
\rotatebox{90}{\hskip 50pt Fringe amplitude}
\hskip -34pt
\begin{minipage}[b]{.9\linewidth}
\vskip -25pt
\centering\epsfig{file=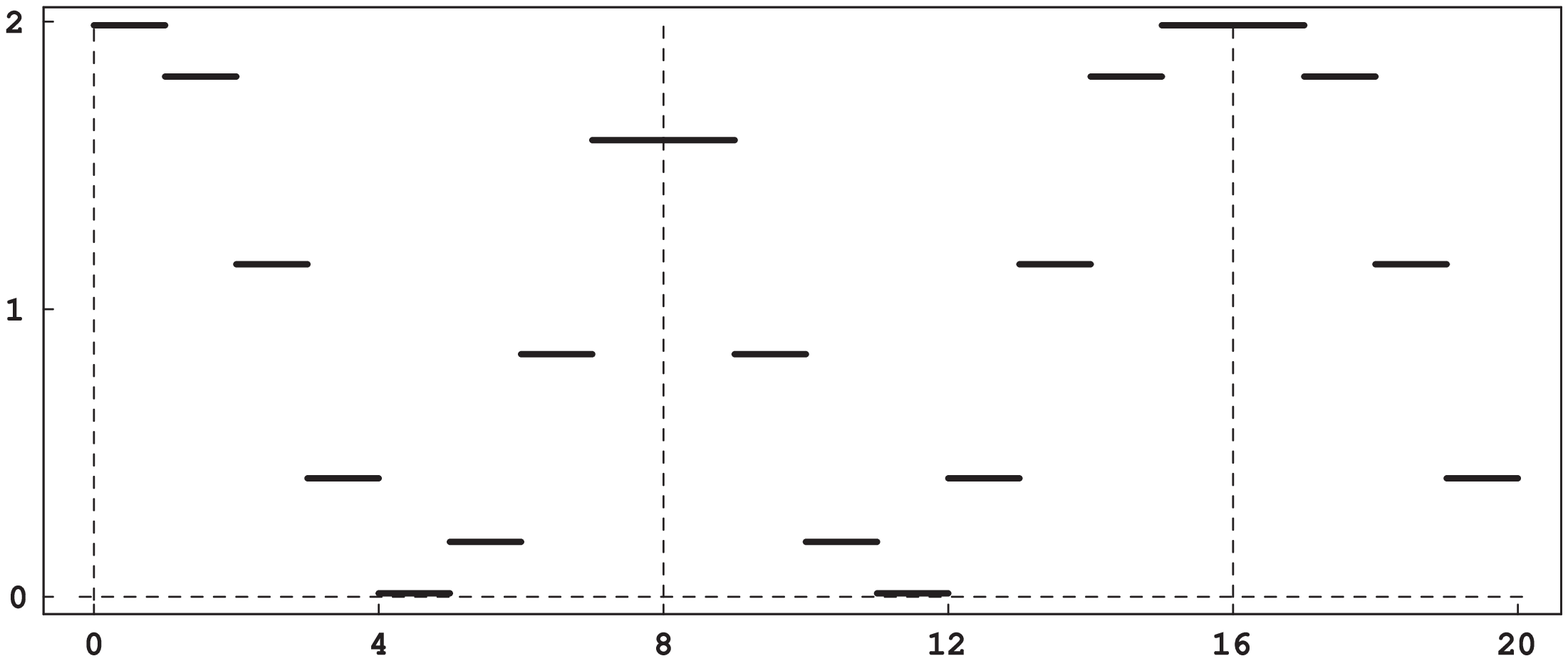,width=11cm}
\vskip -5pt \hskip -0pt
\rotatebox{0}{\hskip 17pt  Optical pathlength difference,~[bins] }
\end{minipage}\vskip -5pt
\caption{Upper plot shows typical OPD modulation stroke with stepping over 8 equal
temporal bins with duration of  10 ms. Lower plot is a monochromatic fringe as a
function of OPD  modulated  in 8 equal steps over a wavelength, as shown in the
upper plot.  (The fringe parameters the same as in Figure \ref{fig:m_fringe}.) 
\label{fig:fringe_step}}
\end{center}
\end{figure}

As shown in the Figures \ref{fig:fringe_ramp} and
\ref{fig:fringe_step} there may be different ways of modulating the internal OPD in an
interferometer.  In particular, Figure \ref{fig:fringe_ramp} demonstrates the case
when the OPD is modulated linearly. Thus, the upper plot
shows a typical phase change with ramping over 8 equal temporal bins with duration of 
10 ms each. The lower plot shows the same monochromatic fringe shown in Figure
\ref{fig:m_fringe} as a function of OPD that is modulated  by ramping over a wavelength.
Note the difference between this case and the case shown in Figure
\ref{fig:fringe_step}, where the upper plot shows a typical phase modulation provided
by the stepping OPD modulation with the stroke stepping over 8 equal temporal bins with
duration of  10 ms each. The lower plot in this Figure demonstrates behavior of
monochromatic fringe as a function of  OPD with modulated stroke stepping  in equal
steps over a wavelength as shown in the upper plot.  

The photon count  on the detector, $N$, is proportional to the intensity,
$N\propto{\cal I}$. Therefore, by collecting photons $N_i(k)\sim{\cal I}(k,t_i)$,
coming at the detector at a certain time intervals $ t_i\in[t_i^-,t_i^+]), ~i
\in [1,..,N]$ (or by integrating Eq. (\ref{eq:int}) over $dt$ from $t_{i-1}$ to $t_i$),
one forms the system of equations to determine the unknown quantities  ${\cal I}_0$, 
$V$ and
$\phi_0$. These time intervals correspond to different values of OPD, $x(t)\in
\Big\{x_i, ~i \in [1,..,N]\Big\}$, therefore,  observational equation in the
case of monochromatic light and ramping OPD  modulation takes the following form:
\begin{equation}
 {\cal I}_i(k) = {\cal I}_0
\Big(1+V  {{\rm sinc}\,[\frac{1}{2} k v \,\Delta\tau_i]}
\sin\big(  \phi_0+ k\,x_i\big)\Big),
\label{eq:N}
\end{equation}
where ${\rm sinc}(\cdot)$-function is given as usual ${\rm sinc} \,z =\sin z/z$; 
$v$ is the constant velocity of OPD modulation stroke,
$x(t)=v t$ and $\Delta\tau_i=t_i-t_{i-1}$ is the integration time for the $i$-th
temporal bin.\footnote{ Note, by taking the limit $v\rightarrow 0$ in Eq.
(\ref{eq:N}) (i.e.
${\sin \!\!c\,[\frac{1}{2} k v \,\Delta\tau_i]} \rightarrow 1$) one recovers the case of
stepping OPD  modulation with a familiar  simple form of observational equation:
${\cal I}_i(k) = {\cal I}_0
\Big(1+V\sin\big(  \phi_0+ k\,x_i\big)\Big).$
}

Eq. (\ref{eq:N}) is the most studied equation when estimating the fringe parameters in
the monochromatic light approximation. Solution to this equation is quite
straightforward and it was extensively discussed in literature (see, for example,
Ref.\citeonline{creath,greivenkamp}). Usually, Eq. (\ref{eq:N}) is represented in a
matrix form as ${\cal I}_i=A_{i\alpha}X^\alpha$, where
indexes $i$ and $\alpha$ running as $i\in\{1,..., N\}$ and $\alpha\in\{1,2,3\}$.
Vector $X^\alpha$ is the to-be-determined phasors vector given as  
$X^\alpha=\big({\cal I}_0;\, {\cal I}_0 V\cos \phi_0;\, 
{\cal I}_0 V\sin\phi_0\big)^T$. Matrix  
${\bf A}^T=A_{i\alpha}=
\Big( 1;\,{{\rm sinc}\,[\frac{1}{2} k v \,\Delta\tau_i]}\sin k x_i; 
\,{{\rm sinc}\,[\frac{1}{2} k v \,\Delta\tau_i]}\cos k x_i\Big)$ is the $3\times N$ 
matrix of 3D rotation in the phase space. A solution to this equation is given by
$X^\alpha= {A^\dagger}^{i\alpha}{\cal I}_i,$   where  ${{\bf A}^\dagger}= ({\bf
A}^T{\bf A})^{-1}{\bf A}^T,$ with  ${{\bf A}^\dagger}$ being the pseudo-inverse of
${\bf A}$.  This set of equations may be solved uniquely only in the case when
$N=3$. For all other cases, when $N>3$, the obtained system of equations is
over-determined, and one obtains a least-squares solution by constructing 
pseudo-inverse matrix 
${{\bf A}^\dagger}$.\cite{Turyshev2000,creath,greivenkamp} We will discuss an
optimally-weighted, noise optimized solution to this set of equations in more details
in Section \ref{sec:pol_g}, while dealing with a more complicated case of the
polychromatic light.

When dealing with a polychromatic light content in the wide bandwidth, one
either i) forms a light with a narrow spectral  width, such that effects of
polychromacity (discussed further) are negligible, or ii) disperses the light beam on a
large number of spectral channels, such that monochromatic approximation is valid
within each channel.\cite{Lawson00} As a result, most of the current algorithms and
simulations for stellar optical interferometry are based on the properties of the
monochromatic light. This is a good approximation for some of existing testbed
configurations that use as many as 80 spectral channels for dispersed light. Nominally
the SIM flight system will use four to eight channels for guide interferometers.
Because of the large bandwidth of each channel (87.5 nm), the quasi-monochromatic
assumptions are not valid, and modifications to the algorithms are
necessary.\cite{markscott,mct} In the following Section we will introduce a method
designed to address this issue. 

\section{Modeling Observables for a Polychromatic Fringe }
\label{sec:model}

\begin{figure}[p]
\noindent  
\begin{center} \hskip  0pt 
\rotatebox{90}{\hskip 70pt Fringe amplitude}
\hskip -10pt
\begin{minipage}[b]{.46\linewidth}
\vskip -10pt
\centering\psfig{figure=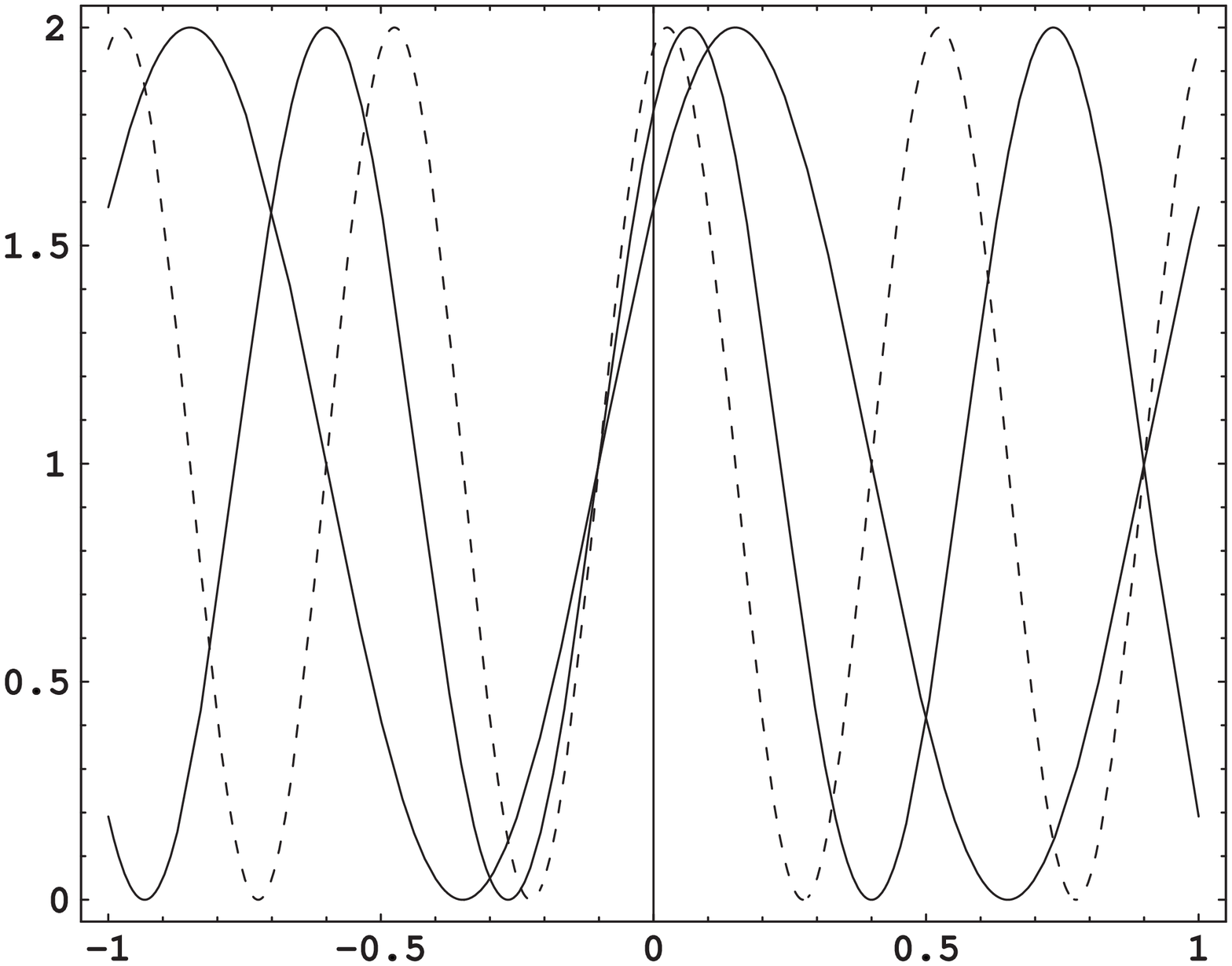,width=65mm,height=60mm}
\rotatebox{0}{\hskip 20pt  Optical pathlength difference,~$x$ }
\end{minipage}
\hskip 10pt
\rotatebox{90}{\hskip 70pt Fringe amplitude}
\hskip -10pt
\begin{minipage}[b]{.46\linewidth}
\vskip -25pt
\centering \psfig{figure=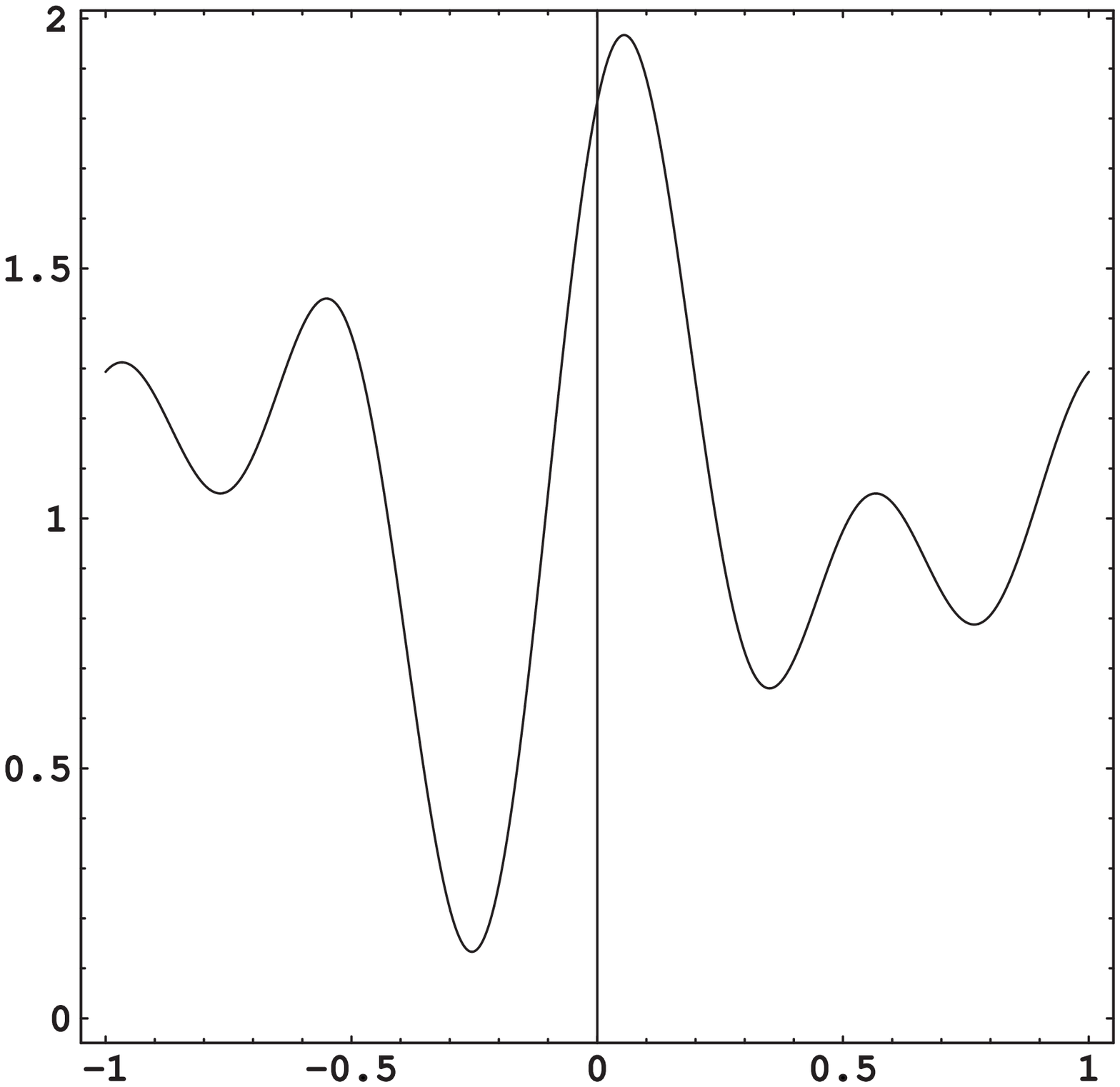,width=65mm,height=60mm}
\rotatebox{0}{\hskip 20pt  Optical pathlength difference,~$x$}
\end{minipage}
     \caption{Left plot shows three independent   monochromatic
waves  presented  as a functions of the OPD. The phase of the fringes is 
changing as $\phi=\phi_0+  k_ix$, with external delay corresponding to a 
phase difference of $\phi_0=\frac{\pi}{5}$ and three wavenumbers chosen $k_1=k_0,~
k_2=1.5k_0, ~k_3=2k_0$ (thus the bandwidth is $\Delta k=k_0$), where $k_0$ is a
reference wavenumber.  The right plot represents the  interference pattern of  
polychromatic light   composed from the same three sinusoidal waves shown on the left
plot. Note the drastic change in the character of the interferometric pattern.
      \label{fig:jupiter_defl}}
 \end{center}
\end{figure}

The observational conditions in the case of polychromatic light are significantly
altered compare to the simplicity of the monochromatic situation discussed in Section
\ref{sec:mono}. Thus, Figure \ref{fig:jupiter_defl} shows a general behavior of
harmonic signals with a different frequencies. The left plot in the figure shows three
sinusoidal monochromatic  waves with the same initial phase offset $\phi_0=\pi/5$. 
Depending on the wavenumber,  each wave, as a function of OPD, produces different
fringe pattern.  Note that a combination of these three waves (a simple model of a
polychromatic fringe) also produces a harmonic signal, but its shape is drastically
different from the initial one (shown on the right plot). For the expected SIM finite
bandwidth the wavenumbers of the interfering light may be different as much as twice
from each other (i.e. the SIM wavenumber bandwidth is $k\in[450,950]$ nm). Thus, in
general, only at zero OPD (i.e. for the white light fringe) the interferometric
pattern would have maximum intensity. 

In this Section we will derive  an equation to describe the interference pattern of a
polychromatic fringe as needed for the SIM CCD detector read-out.

\subsection{Modeling Observables in the Wide Bandwidth Case }

As was mentioned earlier, the case of describing interferometric
pattern that involves a finite bandwidth - is a more complicated one. In
general all the quantities involved are complicated functions of a
wavenumber.  Note that, contrary to the Eq. (\ref{eq:em_mono}), not only the
phase of the electromagnetic wave depend on time and wavelength, but the same true
on it's amplitude  
\begin{equation}
\vec{E}(k,t)=\vec{E}_0(k,t)e^{j\,(\omega t-kx)},
\label{eq:em_mono*}
\end{equation}
\noindent with $\vec{E}_0$ being the wave's amplitude vector that depends  both 
on the wavelength  and time. This expression shows a wave-packet
that describes   individual contribution to the polychromatic light that is
due to a particular monochromatic constituent of it.    

A way to describe this process is to  collect all   constituents of polychromatic
light at different wavelengths that are present in the incoming electromagnetic wave.
It is convenient to express the intensity of the  polychromatic  radiation as a
coherent addition of the individual wave-packets (\ref{eq:em_mono}) with a different
frequencies \cite{goodman}. Denoting 
${\vec E}_{01}$ and ${\vec E}_{02}$,  to represent the light coming onto a
detector from the two arms of  interferometer, this procedure may
functionally be expressed in the following form:
\begin{eqnarray}
{\cal I}(k,t)&\sim& \Big\langle
\big({\vec E}_{01}(k,t)+{\vec E}_{02}(k,t)\big)\cdot
\big({\vec E}^*_{01}(k,t)+{\vec E}^*_{02}(k,t)\big)
\Big\rangle_{\sf time}= \nonumber\\ 
&=&{\vec E}^2_{01}(k)+{\vec E}^2_{02}(k) +
2|{\vec E}_{01}(k)||{\vec E}_{02}(k)|\gamma(k)
\cos\Big(\Phi(k)+\Delta\varphi_{12}(k,t)\Big),
\label{eq:beg}
\end{eqnarray}

\noindent where $\Delta\phi_{12}(k,t)$ is the external phase difference between 
the two arms of interferometer and $ \tilde \gamma(k) =\tilde {\cal V}(k){\tilde
T}(k)=\gamma(k)  e^{j\,\Phi(k)}$ is the  complex coherency factor. 

This complex coherency factor includes both - the  true  complex
visibility of the source, $\tilde {\cal V}(k)$, and the complex instrumental transfer
function denoted as ${\tilde T}(k)$. The complex source visibility function is given
in its  usual form  
\begin{equation}
\tilde {\cal V}(k)={\cal V}(k) e^{j\,\phi_{\cal V}(k)},
\label{eq:v0}
\end{equation}
with ${\cal V}(k)$ being the amplitude and $\phi_{\cal V}(k)$ the phase of
the true source visibility.  According to Van-Cittert-Zernike theorem,
namely this quantity is connected to the true radiation emitted by a star.
Specifically,  this theorem allows to calculate mutual intensity observed on a
surface some distance from the source  \cite{goodman}. 
The instrumental complex transfer
function, ${\tilde T}(k)$,   may also be presented in a similar manner
\begin{equation}
{\tilde T}(k)=T(k) e^{j\,\phi_T(k)},
\label{eq:mu0}
\end{equation}
with $T(k)$ being the amplitude of this function and $\phi_T(k)$ its phase.
 Therefore, the complex coherency factor ${\tilde \gamma}(k)$,  that is present in
the equation (\ref{eq:beg}), has  the following functional dependency:
\begin{equation}
\tilde \gamma(k)=\tilde {\cal V}(k){\tilde T}(k)=
{\cal V}(k) T(k) e^{j\,[\phi_{\cal V}(k)+\phi_T(k)]}\equiv
\gamma(k)  e^{j\,\Phi(k)},
\label{eq:gam}
\end{equation} 

\noindent where combined apparent visibility and phase are given by the expressions
$\gamma(k)= {\cal V}(k){T}(k)$ and  $\Phi(k)= \phi_{\cal
V}(k)+\phi_T(k)$ correspondingly.

The  modulated phase difference $\Delta\varphi_{12}(k,t)$ in the
Eq.(\ref{eq:beg}) may be expressed in terms of the internal pathlength difference:
\begin{equation}
\Delta\varphi_{12}(k,t)=kx(t),
\label{eq:pat}
\end{equation}  
\noindent with $x(t)=\ell_1(t)-\ell_2(t)$ being the internal pathlength difference
between the two arms of  interferometer (see Figure \ref{fig:astrom}). 

Let us  define  individual spectral densities of photon flux in each arm of 
interferometer as 
${\cal I}_1\sim {\vec E}^2_{01}(k)$ and ${\cal I}_2\sim {\vec E}^2_{02}(k)$.   
This allows us to present the  intensity of  incoming radiation at a particular
wavenumber $k$ as  
\begin{equation}
{\cal I}(k,t)\sim{\cal I}_{1}(k)+{\cal I}_{2}(k)+
2\sqrt{{\cal I}_{1}(k){\cal I}_{2}(k)}\,
\gamma(k)\cos\big(\Phi(k)+kx(t)\big).
\label{eq:dinen}
\end{equation} 

\noindent It is naturally to define the total spectral density of photon flux of
light approaching the detector  as ${\cal I}_0(k)={\cal I}_{1}(k)+{\cal I}_{2}(k)$.
Note that  individual spectral densities of photon flux in the two arms of  
interferometer,
${\cal I}_{1}$ and ${\cal I}_{2}$, may be different. To account for such a
mismatch we introduce apparent visibility ${\tilde V}(k)$  
\begin{equation}
{\tilde V}(k)=\frac{2\sqrt{{\cal I}_{1}(k){\cal I}_{2}(k)}}
{{\cal I}_{1}(k)+{\cal I}_{2}(k)} \gamma(k) =V(k)e^{j\,\Phi(k)}.
\label{eq:ap_vis}
\end{equation}
 
\noindent The resulted expression for the spectral density of photon flux  of the
incoming radiation at a particular  wavenumber $k$  takes the following form:
\begin{equation}
{\cal I}(k,t)= {\cal I}_0(k)
\Big(1+V(k)\cos\big(\Phi(k)+kx(t)\big)\Big).
\label{eq:din}
\end{equation} 

Furthermore, the energy density of incoming polychromatic radiation per a unit
wavenumber may be obtained by integrating the photon flux at the detector Eq.
(\ref{eq:din}) over the wavenumber space $dk$:
\begin{equation}
d{\cal I}(k,t)= {\cal I}_0(k)
\Big(1+V(k)\cos\big(\Phi(k)+kx(t)\big)\Big)dk.
\label{eq:din1}
\end{equation} 

\noindent Expression (\ref{eq:din1}) defines the energy density  of  the
interfering light per a unit wavenumber.  The 
quantity that is proportional to it, namely the photo-electron count, will be
registered by a CCD detector via a photo-electron emission process and will be
discussed in  Section \ref{sec:model}\ref{sec:photo_e}.

\subsection{Effect of a Beam Splitter}
\label{sec:beamsplitter}

To complete the formulation of our model we need to account for yet one more
element that is of crucial importance for a Michelson stellar interferometer -
the beam splitter.\cite{colavita1,shao2} It is well-known that one of the figures of
merit when considering various beam splitter designs for the astrometric beam
combiner, is the variation of phase as a function of wavelength between the interfering
beams at zero OPD.  Thus, ideally, one would like zero phase difference for
all wavelengths at zero OPD since this would produce a fringe maximum
simultaneously at all wavelengths (or fringe minimum for a Michelson stellar
interferometer).  Another desirable case is one in which the phase difference varies
linearly with frequency.  In this case one can still achieve a simultaneous fringe
maximum at all wavelengths; however, for this case there is a constant correction that
must be applied to the calculated external delay. \cite{Turyshev2000} 
 
This effect is important because it may produce a significant
contribution to the OPD. We may include this effect in a most general way ---
the wavenumber dependent function $f(k)$ added to the desirable $\pi/2$ effect
of an ideal beam splitter. Thus, the total effect of the beam splitter may be
modeled as
\begin{equation}
\delta \Phi_{\sf bs}(k)=\frac{\pi}{2} + f(k),
\label{eq:bsp}
\end{equation}
where $f(k)$ is a slow varying function of a wavenumber. This function may be
approximated up to the second order around some central frequency  of bandpass
(the whole 80 channels pass), $k_c,$  as
\begin{equation}
f(k)=f(k_c)+f'(k_c)(k-k_c)+ \frac{1}{2}f''(k_c)(k-k_c)^2+
{\cal O}(\Delta k^3_c).
\end{equation}
It may be shown\cite{Turyshev2000} that  function  $f(k_c)$
will have a direct impact on the phase accuracy estimation by shifting the phase by a
constant value.  In addition, the function $f'(k_c)$ will change the envelope function
and, in general, will produce a non-linear contribution to the phase. $f''(k_c)$ will
have impact on both - linear contribution to the phase and the non-linear one;  first
it comes as a correction to the envelope function and then to the phase. Even though,
the magnitude of the effects of $f''$ is smallest among all,   the parameters $f, f',
f''$ are all of importance. The corresponding effects may be well modeled, depending
on the optical properties of the beam splitter, but the final answer would come
probably form calibration.

Finally, without loosing generality, the effect of 
beam splitter may be accounted for as an additional phase shift to the argument 
besides the usual $\delta \Phi_{\sf bs}(k)=\frac{\pi}{2}$ and, thus leading to 
a new definition for the fringe phase $\phi(k)$ in the form  
\begin{equation}
\Phi(k)\qquad \rightarrow \qquad \Phi(k)+\delta \Phi_{\sf bs}(k)
=\Phi(k)+\frac{\pi}{2}+f(k)=\frac{\pi}{2}+\phi(k).
\label{eq:pha*}
\end{equation}
Hence,  the light intensity Eq. (\ref{eq:din1}) may be given as follows:
\begin{equation}
d{\cal I}(k,t)= {\cal I}_0(k)
\Big(1+V(k)\sin\big(\phi(k)+kx(t)\big)\Big)dk,
\label{eq:din_bs}
\end{equation} 
\noindent where we accounted for the nominal $\frac{\pi}{2}$ phase shift due to
the beam splitter.

\subsection{Detector's Photo-electron Counts}
\label{sec:photo_e}

A CCD detector is responding a quantity that is closely related to the intensity of
radiation,  namely the incoming energy which is given as 
$d{\cal E}(k,t)=d{\cal I}(k,t)dt$, or 
\begin{equation}
d{\cal E}(k,t)= {\cal F}_0(k){\cal I}_0(k)\Big(1+V(k)
\sin\big(\phi(k)+kx(t)\big)\Big)dkdt,
\label{eq:det_ener}
\end{equation}
where $ {\cal F}_0(k) $ is a dimensionless factor representing 
the total instrumental throughput.\cite{markscott,mct}
Usually a detector is optimized to work at a certain wavelengths better then at the
others. This property may be qualitatively described by, so called, the quantum
efficiency of a detector, $\alpha(k)$. Thus, the quantum efficiency of a CCD
detector is conventionally defined as  ${\cal N}(k,t)= \alpha(k) {\cal E}(k,t)$,
therefore the density of the emitted photo-electrons per a unit area may be 
presented by the following expression:
\begin{equation}
d{\cal N} (k,t) =  \alpha(k) {\cal F}_0(k){\cal I}_0(k)\Big(1+V(k)\sin
\big(\phi(k)+kx(t)\big)\Big)dk\,dt.
\label{eq:det_alp}
\end{equation}
\noindent Note that quantum efficiency of the detector   is a  function of a
wavelength $\alpha(k)$. This dependency  will not addressed in the present study, but
will be explored elsewhere.

The total number of photo-electron counts, $N$, depends on the collective
area of the detector in accord $dN(k,t)={\cal N}(k,t)dA$. Note that the total
photon count is a function of power of  radiation approaching the detector  which
is given as  $d{\cal P}(k,t)={\cal E}(k,t)dA$.  Therefore, one obtains 
\begin{equation}
dN(k,t) =  \alpha(k) {\cal F}_0(k){\cal I}_0(k)\Big(1+V(k)\sin
\big(\phi(k)+kx(t)\big)\Big)(\vec{n}\cdot\vec{\tau_A}) dAdk\,dt,
\label{eq:det_al*}
\end{equation}
\noindent where $\vec{n}$ is the direction of the wave falling on  the detector and 
$\vec{dA}=dA\vec{\tau_A}$ is the vector of the collective area of the detector.
Integration of this equation over the collective area depends on the properties of
the experimental setup, notably on the orientation of the collective area with respect
to incoming light. Thus, a small geometric misalignment in the system may be
responsible for the uneven illumination of different pixels on the detector.
Additionally, the wavefront tilt may produce a measurable contribution to the
fringe parameters estimated with the help of Eq. (\ref{eq:det_al*}). 

In Section \ref{sec:filter} we will introduce a concept of a filtered polychromatic
light for which notation $dA_\ell$ will be designated for the collective area of the
${\ell}$-th spectral channel (or area of illumination on the detector designated for a
particular spectral channel).  Assuming that this area  is small, and neglecting 
divergence of the radiation in the instrument, we may integrate Eq. (\ref{eq:det_al*})
over $dA_\ell$. Note, that this operation, together with the quantum efficiency of
the detector,  mathematically may be ``folded'' into the definition for the
bandpass filter (also discussed Section \ref{sec:filter}). Therefore,  we can 
designate to the filter not only the function of control over the  allowed band-pass
of incoming  radiation, but also it can produce masking of the detector by allowing
exposure of only a certain areas of it.  Note that we ignore the effects
of the wavefront tilt and uneven illumination of different pixels on the detector, and
defer the discussion of these issues to a subsequent publication. 

Assumptions above result in the  following expression for
the density of the photon-counts  registered by the detector:
\begin{equation}
N(k,t)={\cal F}(k){\cal I}_0(k)\Big(1+V(k)\sin\big(\phi(k)+kx(t)\big)\Big),
\label{eq:n0}
\end{equation}
where we corrected the total instrumental throughput for the quantum efficiency 
of the detector, ${\cal F}(k)=\alpha(k){\cal F}_0(k)$. (This form of the
notation is quite sufficient for the error propagation and sensitivity
analyses that we will report elsewhere.) Therefore, we derived a model for the
interferometric fringe pattern that accounts for a number of effects of light
propagating through the instrument.  In the next Section we will derive observational
equation to be used for estimating the fringe parameters if interest. 


\section{Parameterization of a Polychromatic Fringe Pattern}
\label{sec:integr}

As we see from the previous discussion, description of the interferometric
pattern in the  polychromatic case that involves a finite bandwidth of radiation - is
a technically complicated task. Thus, the observational conditions in the case of
polychromatic light are significantly altered compare to the simplicity of the
monochromatic process.  In general, all the quantities involved are complicated
functions of the  wavelength. In the previous Sections, we choose a way to describe
this process is to  collect contributions of all infinitesimal  constituents of
polychromatic light at different wavelengths  within the bandwidth of the incoming
electromagnetic radiation.\cite{Turyshev2000} 

The total number of photo-electron counts, $N$,  registered by a CCD
detector per wavenumber and per unit time, may be given by the following
expression:
\begin{equation}
dN (k,t) =  {\cal F} (k){\cal I}_0(k)\Big(1+V(k)\sin
\big[\phi(k)+kx(t)\big]\Big) dk\,dt,
\label{eq:photon_count}
\end{equation}

\noindent where $ {\cal F}(k) $ is a dimensionless factor representing 
the total instrumental throughput;\cite{markscott,mct}
${\cal I}_0(k)$, $V(k)$ and $\phi(k)$ are the spectral density of photon flux,
visibility and phase of the  incoming light; $x(t)$ is modulated internal delay. We
are using a nomenclature where a wavenumber $k$  relates  to the wavelength as follows
$k=\frac{2\pi}{\lambda}$. We also accounted for the nominal $\frac{\pi}{2}$ phase
shift due to the SIM beam splitter, which produces a sine fringe rather than a cosine
one.   

Note that  the total instrumental throughput depends on a
number of other factors, some of these are the collective area of the
detector, quantum efficiency of  CCD, and overall spectral response of the
instrument (as discussed in  Section \ref{sec:model}\ref{sec:photo_e}).  Our goal  here
is to derive observational equation that may be used to estimate the apparent fringe
phase and visibility.  To estimate the true source visibility  and phase one would
have to perform a set of additional calibration and estimation procedures that will be 
addressed elsewhere.

\subsection{Integration Over the  Spectral Bandwidth}

In this Section we will perform integrations of  Eq. (\ref{eq:photon_count}) over
wavenumber space and time,  that are necessary to derive analytical model. This
model will be used further for the purposes of the fringe parameters estimation. 

Let us first perform   integration over the SIM wavenumber bandwidth
$k\in[k_{\tt SIM}^-,k_{\tt SIM}^+]$, where $k_{\tt SIM}^-=450 $ nm is the beginning
of the SIM bandwidth, and  $k_{\tt SIM}^+=950$ nm is the end of this bandwidth,
thus $k\in[450,950]$ nm. A formal integration of 
Eq.(\ref{eq:photon_count}) over $dk$ leads to the following result 
\begin{eqnarray}
N(t){\Delta k_{\tt SIM}}&=&  
\int_{k_{\tt SIM}^-}^{k_{\tt SIM}^+} \!\! N (k,t)
\,dk, \qquad\qquad \Delta k_{\tt SIM}=
k_{\tt SIM}^+-k_{\tt SIM}^-.
\label{eq:Ndk}
\end{eqnarray}

\noindent In the case of channeled (or dispersed)
spectrum output, the integration of this equation  over the range of wavenumbers
is straightforward.  For this purpose, we designate index,
$\ell$,   to denote a particular spectral channel. Suppose that there exists a total
of $L$ spectral channels, thus $\ell\in[1, ..., L]$.  

Our definition for the spectral channel $\ell$ implies the width of the channel 
$\Delta k_\ell=k_\ell^+-k_\ell^-$ and existence of a ``central'' wavenumber
$k_\ell$ within this channel. We also, assume continuous spectrum within the   
bandwidth, so that there is no gaps exist in the interval
$k\in[k_{\tt SIM}^-,k_{\tt SIM}^+]$. A consequence of this is the equality
$k_{\ell+1}^-=k_{\ell}^+$, which leads to the following discrete representation of
the bandwidth  
$
\Delta k_{\tt SIM}=
\sum_{\ell=1}^L\, \Delta k_\ell.
$
These assumptions allow us to rewrite the right-hand side of Eq.
(\ref{eq:Ndk}) as follows:
\begin{equation}
\int_{k_{\tt SIM}^-}^{k_{\tt SIM}^+} \!\! N (k,t)\,dk=
\sum_{\ell=1}^L\,\int_{k_{\ell}^-}^{k_{\ell}^+} \!\! N (k,t)
\,dk = \sum_{\ell=1}^L\,
N_\ell(t) \Delta k_{\ell}, 
\label{eq:Ndk0}
\end{equation}
\noindent where $N_\ell(t) $ is the
instantaneous number of photons within a particular spectral channel and has the
following form 
\begin{eqnarray}
N_\ell(t)= \frac{1}{\Delta k_{\ell}}
\int_{k^{-}_\ell}^{k^{+}_\ell} \!\!  {\cal F} (k){\cal I}_0(k)\Big(1+V(k)\sin
\big[\phi(k)+kx(t)\big]\Big) dk.
\label{eq:Nell}
\end{eqnarray}
This equation is our first important  result. It will help to focus our
attention from the discussion of coherent processes within the whole wide bandwidth, 
onto addressing this processes on a smaller scale -- within a particular 
narrow spectral channel, $\ell$.

\subsection{Definitions for the Fringe Parameters}
\label{sec:notations}
 
At this point, it is convenient to introduce a set of useful notations. 
First of all, we define the average total intensity of incoming electromagnetic
radiation, ${\cal I}_{0 \ell}$, within the ${\ell}$-th spectral channel as
\begin{eqnarray}
{\cal I}_{0 \ell}=\frac{1}{\Delta k_\ell}
\int_{k^{-}_\ell}^{k^{+}_\ell} \!\!{\cal F}(k){\cal I}_0(k)dk.
\label{eq:Ij}
\end{eqnarray}
It is natural to introduce normalized intensity of light
$\hat{{\cal I}}_{0\ell}$  within the  ${\ell}$-th channel: 
\begin{equation}
\hat{{\cal I}}_{0\ell}(k)=
\frac{{\cal F} (k){\cal I}_0(k)}{{\cal I}_{0 \ell}}  
\qquad\qquad {\rm with } \qquad\qquad
\frac{1}{\Delta k_\ell}
\int_{k^{-}_\ell}^{k^{+}_\ell} \!\!\hat{{\cal I}}_{0\ell}(k)dk=1.
\label{eq:Ijnorm}
\end{equation}

\noindent These new notations allow  to present Eq. (\ref{eq:Nell}) as given below
\begin{equation}
N_\ell(t) =  {\cal I}_{0 \ell}\bigg(1 + \frac{1}{\Delta k_\ell}
  \int_{k^{-}_\ell}^{k^{+}_\ell} \!\!\hat{\cal I}_{0 \ell}(k) V(k)
\sin\big[\phi(k)+kx(t)\big]dk \bigg).
\label{eq:bigNt}
\end{equation}
In the next Section we will define the fringe visibility, phase and mean
wavenumber.

\subsection{Fringe Visibility, Mean Wavenumber and Phase}

To further simplify the obtained equation, we will introduce functional form the
fringe visibility, the phase and the wavenumber notations. Thus, the fringe
visibility,
$V_{0\ell}$, within the ${\ell}$-th channel is given as 
\begin{eqnarray}  
V_{0\ell}=\frac{1}{\Delta k_\ell}
\int_{k^{-}_\ell}^{k^{+}_\ell} \!\! \hat{\cal I}_{0 \ell}(k)V(k)dk.
\label{eq:v0j}
\end{eqnarray}

\noindent Similarly to Eq.~(\ref{eq:Ijnorm}) we denote  normalized
visibility in the channel as
\begin{equation}
\hat{V}_{0\ell}(k)=\frac{\hat{\cal I}_{0 \ell}(k)V(k)}{V_{0\ell}}\equiv
\frac{{\cal F} (k){\cal I}_0(k)V(k)}{{\cal I}_{0 \ell}V_{0\ell}},
\qquad\qquad
\frac{1}{\Delta k_\ell}
\int_{k^{-}_\ell}^{k^{+}_\ell} \!\!\hat{V}_{0\ell}(k)dk=1.
\label{eq:v0jno}
\end{equation}
These definitions help us to re-write equation (\ref{eq:bigNt}) in the
following compact form
\begin{equation}
N_\ell(t)=  {\cal I}_{0 \ell}\Big( 1+
V_{0\ell}\frac{1}{\Delta k_\ell}
\int_{k^{-}_\ell}^{k^{+}_\ell}
\!\!\hat{V}_{0\ell}(k)
\sin\big[\phi(k)+kx(t)\big]dk\Big).
\label{eq:nj}
\end{equation}



To define  mean wavenumber, $k_\ell$, and mean
phase, $\phi_\ell$, for the ${\ell}$-th spectral channel we will use the 
following expression:
\begin{eqnarray}
k_\ell&=&\frac{1}{\Delta k_\ell}\int_{k^{-}_\ell}^{k^{+}_\ell} \!\!
\hat{\cal I}_{0\ell}(k)\,k\,dk ~\equiv~
\frac{1}{{\cal I}_{0 \ell}\Delta k_{\ell}}
\int_{k^{-}_\ell}^{k^{+}_\ell} \!\!
{\cal F} (k){\cal I}_0(k)\,k\,dk, 
\label{eq:kl}
\end{eqnarray}
There are two ways to define the phase within the channel. 
Thus, it is tempting to define the mean phase as 
\begin{equation}
\phi_\ell=\frac{1}{\Delta k_\ell}
\int_{k^{-}_\ell}^{k^{+}_\ell} \!\!\hat{\cal I}_{0\ell}(k)\phi(k)dk 
~\equiv~
\frac{1}{{\cal I}_{0 \ell}\Delta k_{\ell}}
\int_{k^{-}_\ell}^{k^{+}_\ell} \!\!
{\cal F} (k){\cal I}_0(k)\phi(k)dk.
\label{eq:fj0}
\end{equation}
This definition is acceptable for  narrow spectral channel, however for a wide
channel one needs  a more convenient form, namely 
\begin{equation}
\phi(k_\ell), \qquad\qquad {\rm ~which ~is } 
\qquad\qquad \phi(k_\ell)\not=\phi_\ell,
\label{eq:ph}
\end{equation}
and  is simply the phase value at the central wavenumber. In our further
analysis we will be using this latter definition. (The relationships between the two
definitions for the phase Eqs. (\ref{eq:fj0}) and (\ref{eq:ph}) will  be addressed in
Appendix \ref{sec:app_phase}). 

The three introduced quantities (i.e. visibility, mean wavenumber $k_\ell$ and  
phase at the mean wavenumber $\phi(k_\ell)$) allow  to proceed with integration of Eq.
(\ref{eq:nj}).

\subsection{Complex Fringe Envelope Function}


Definitions introduced in the previous Section allow us to separate
functions $k_\ell$ and $\phi(k_\ell)$ from the functions with direct dependency on the
wavenumber $k$. As a result, Eq. (\ref{eq:nj}), for the total photon count, may be
presented as below
\small
\begin{eqnarray}
N_\ell(t)&=& {\cal I}_{0 \ell}\bigg(1 \,+  
\nonumber\\[0pt]
&&\hskip -35pt\,+\,\,
V_{0\ell}\sin\big[\phi(k_\ell)+k_\ell x(t)\big]\,
\frac{1}{\Delta k_\ell}
\int_{k^{-}_\ell}^{k^{+}_\ell}
\!\!\hat{V}_{0\ell}(k)
\cos\big[(k-k_\ell)x(t)+\phi(k)-\phi(k_\ell)\big]dk+ 
\nonumber\\[0pt]
&&\hskip -35pt\,+\,\,
  V_{0\ell}\cos\big[\phi(k_\ell)+k_\ell x(t)\big]\,
\frac{1}{\Delta k_\ell}
\int_{k^{-}_\ell}^{k^{+}_\ell}
\!\!\hat{V}_{0\ell}(k)
\sin\big[(k-k_\ell)x(t)+\phi(k)-\phi(k_\ell)\big]dk\bigg). 
\hskip 20pt 
\label{eq:n0jq}
\end{eqnarray} 

To further simplify the analysis, it is convenient to introduce
the complex  fringe envelope function,  
${\tilde W_\ell}\big[\Delta k_\ell, \phi(k_\ell),
x(t)\big]$, which is given as 
\begin{eqnarray}
{\tilde W_\ell}\big[\Delta k_\ell, \phi(k_\ell), x(t)\big]&=&
\frac{1}{\Delta k_\ell}\int_{k^{-}_\ell}^{k^{+}_\ell} \!\!
\hat{V}_{0\ell}(k)\,
e^{j\,\big((k-k_\ell)x(t)+\phi(k)-\phi(k_\ell)\big)}dk \equiv
\label{eq:wj}\\
&\equiv&
 \frac{1}{{\cal I}_{0 \ell}V_{0\ell} 
\Delta k_\ell}\int_{k^{-}_\ell}^{k^{+}_\ell} \!\!
{\cal F}(k){\cal I}_0(k)V(k)\,
e^{j\,\big((k-k_\ell)x(t)+\phi(k)-\phi(k_\ell)\big)} dk.
\hskip20pt
\label{eq:wj?}
\end{eqnarray} 

As a complex function,  ${\tilde W_\ell}$ (please refer to discussion of the fringe
envelope function given in Appendix \ref{sec:appa}) may be equivalently presented by
its real, ${\sf Re}\big\{{\tilde W_\ell}\big\}$,  and imaginary, 
${\sf Im}\big\{{\tilde W_\ell}\big\}$, components:
\begin{eqnarray}
{\tilde W_\ell}&=&
{\sf Re}\big\{{\tilde W_\ell}\big\}+j\,\, 
{\sf Im}\big\{{\tilde W_\ell}\big\}\label{eq:wjll}
\end{eqnarray}
with
\begin{eqnarray}
{\sf Re}\big\{{\tilde W_\ell}\big\}&=&
\frac{1}{\Delta k_\ell}
\int_{k^{-}_\ell}^{k^{+}_\ell}
\!\!\hat{V}_{0\ell}(k)
\cos\big[(k-k_\ell)x(t)+\phi(k)-\phi(k_\ell)\big]dk, 
\nonumber\\[4pt]
{\sf Im}\big\{{\tilde W_\ell}\big\}&=&
\frac{1}{\Delta k_\ell}
\int_{k^{-}_\ell}^{k^{+}_\ell}
\!\!\hat{V}_{0\ell}(k)\,
\sin\big[(k-k_\ell)x(t)+\phi(k)-\phi(k_\ell)\big]dk. 
\hskip 20pt 
\label{eq:n0jqh}
\end{eqnarray} 
 
This definition of complex envelope function given by
Eq.(\ref{eq:wj})-(\ref{eq:n0jqh})  allows us to present expression (\ref{eq:n0jq})  in
a simpler form: 
\begin{eqnarray}
N_\ell(t)&=& {\cal I}_{0 \ell}\bigg(1 \,+  \,
V_{0\ell}\sin\big[\phi(k_\ell)+k_\ell x(t)\big]\,
{\sf Re}\Big\{{\tilde W_\ell}\big[  x(t)\big]\Big\}+ 
\nonumber\\[0pt]
&&\hskip 32pt\,+\,\, V_{0\ell}\cos\big[\phi(k_\ell)+k_\ell x(t)\big]\,
{\sf Im}\Big\{{\tilde W_\ell}\big[  x(t)\big]\Big\}\bigg). 
\hskip 20pt 
\label{eq:n0j}
\end{eqnarray} 
The complex fringe envelope function, ${\tilde W_\ell}\big[ x(t)\big]$, as
any complex function, may also be represented by its amplitude and its phase, namely: 
\begin{eqnarray}
{\tilde W_\ell}\big[\Delta k_\ell, \phi(k_\ell), x(t)\big]&=&
{\cal E}_{\ell}\big[\Delta k_\ell, \phi(k_\ell), x(t)\big]\,
e^{j\Omega_{\ell} [\Delta k_\ell, \phi(k_\ell), x(t)]},
\label{eq:wjs11}
\end{eqnarray}
\noindent where ${\cal E}_{\ell}$ and $\Omega_{\ell}$ are the amplitude and  
phase correspondingly. For the complex
envelope function Eq. (\ref{eq:wjll}) these two are given as follows:
\begin{eqnarray}
{\cal E}_{\ell}(t)&=&
 \sqrt{{\sf Re}^2\big\{{\tilde W_\ell}\big\}+ 
{\sf Im}^2\big\{{\tilde W_\ell}\big\}},\qquad\qquad
\Omega_{\ell}(t)=
{\sf ArcTan}\Big\{
\frac{{\sf Im}\big\{{\tilde W_\ell}\big\}}
{{\sf Re}\big\{{\tilde W_\ell}\big\}}\Big\}. 
\label{eq:wjlls}
\end{eqnarray}
 
Finally, we re-write  Eq. (\ref{eq:n0j}) in the 
following general form:
\begin{eqnarray}
N_\ell(t)&=& {\cal I}_{0 \ell}\bigg(1 +  
V_{0\ell}\,\,{\cal E}_{\ell}(t)
\,\sin\big[\phi(k_\ell)+k_\ell x(t)+
\Omega_{\ell}(t)\big]\bigg). 
\hskip 20pt
\label{eq:n0jj}
\end{eqnarray} 

  Note that the apparent visibility of the fringe now is the
product of the true averaged visibility and the modulus of the Fourier transform 
of the filter function, evaluated at the current delay or
\begin{equation}
\tilde{\Gamma}_x =\tilde{V}_{0\ell}\tilde{W}_{\ell}\equiv
 V_{0\ell}\,{\cal E}_{\ell}\, 
e^{j\big(\phi(k_\ell)+\Omega_{\ell}\big)}.
\end{equation}
\noindent It is known that the transfer function ${\tilde W_\ell}$ describes the 
coherence envelope.\cite{goodman} If $\hat{V}_{0\ell}(k)\sim  {\cal F}(k){\cal
I}_0(k)V(k)$ is symmetric, then ${\tilde W_\ell}$ is real valued,
$\Omega_{\ell}=0$, and only at zero delay,\cite{Lawson00} where the envelope is at
peak, is the true visibility observed.

\subsection{Temporal Integration }
\label{sec:temp_int}

The last integration  to be performed in Eq.(\ref{eq:photon_count}) (or
equivalently Eq.(\ref{eq:n0jj})), is the integration over time. 
The optical pathlength difference may be modulated either as a set of discrete values 
corresponding to a number of steps in the OPD space (stepping OPD modulation) or by
ramping OPD over the range of values.    The total integration time, $\Delta t$,
is the sum of durations of eight temporal bins. (While our result is applicable for
arbitrary number of temporal bins, the SIM design will utilize 8 temporal bins):
\begin{equation}
\Delta t=t^+-t^-=\sum_{i=1}^{N=8}\Delta \tau_i, ~~~~\text{with }~~~~
\Delta \tau_i=t^{+}_i-t^{-}_i.
\end{equation}
Direct integration of  Eq. (\ref{eq:n0j})  leads to  expression for the total
number of photons collected at each stroke of OPD modulation:
\begin{eqnarray}
N_\ell\,\Delta t&=& \int_{t^{-}}^{t^{+}} \!\!
N_\ell(t)
\,dt=\sum_{i=1}^{N=8}\int_{t^{-}_i}^{t^{+}_i} \!\!
N_\ell(t)
\,dt=\sum_{i=1}^{N=8} N_{\ell i}\Delta \tau_i,
\label{eq:Nijint}
\end{eqnarray}
\noindent where $N_{\ell i} \Delta \tau_i$ is the total number of photons
collected in a particular $i$-th temporal bin and for the $\ell$-th spectral channel.
Substituting $N_\ell(t)$ from  Eq. (\ref{eq:n0jj}) directly into 
Eq. (\ref{eq:Nijint}), one obtains following  expression for $N_{\ell i}$:
\small
\begin{eqnarray}
N_{\ell i}  &=&  \frac{1}{\Delta\tau_i}
\int_{t^{-}_i}^{t^{+}_i} \!\!
N_\ell(t)
\,dt = \nonumber\\[6pt]
&=& \frac{1}{\Delta\tau_i}\int_{t^{-}_i}^{t^{+}_i} \!\!
{\cal I}_{0 \ell}\bigg(1 \,+ V_{0\ell}\sin\phi(k_\ell)\,  
{\cal E}_{\ell}(t)
\,\cos\big[k_\ell x(t)+
\Omega_{\ell}(t)\big]+
\nonumber\\[6pt]
&&\hskip 73pt + \,
V_{0\ell}\cos\phi(k_\ell)\, 
{\cal E}_{\ell}(t)
\,\sin\big[k_\ell x(t)+\Omega_{\ell}(t)\big]\bigg)dt. 
\hskip 20pt 
\label{eq:n0jnew}
\end{eqnarray} 
\normalsize
To complete this integration,  we assume that   quantities 
${\cal I}_{0 \ell},$
$V_{0\ell}, \phi(k_\ell)$, and $k_\ell$ do not  change with  time during the
photon-counting intervals. The only quantity that is explicitly varies with time --
is the optical pathlength difference $x(t)$.  

\begin{figure}[t]
\noindent  
\begin{center} \hskip -40pt 
\rotatebox{90}{\hskip 70pt Fringe amplitude}
\hskip -5pt
\begin{minipage}[b]{.46\linewidth}
\vskip -25pt
\centering\psfig{figure=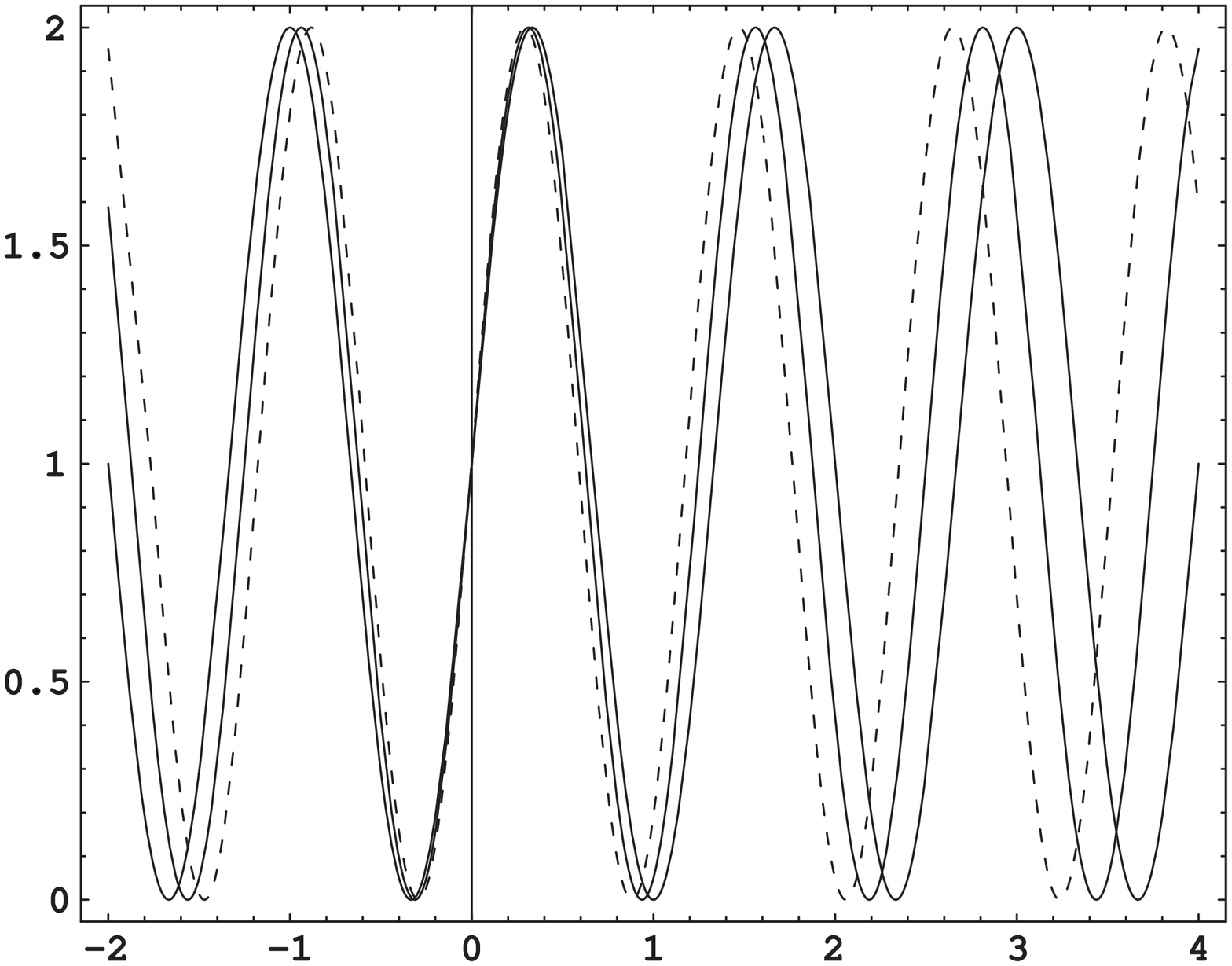,width=72mm,height=60mm}
\rotatebox{0}{\hskip 18pt  Optical pathlength difference,~$x$ }
\end{minipage}
\hskip 10pt
\rotatebox{90}{\hskip 70pt Fringe amplitude}
\hskip -50pt
\begin{minipage}[b]{.46\linewidth}
\vskip -25pt
\centering \psfig{figure=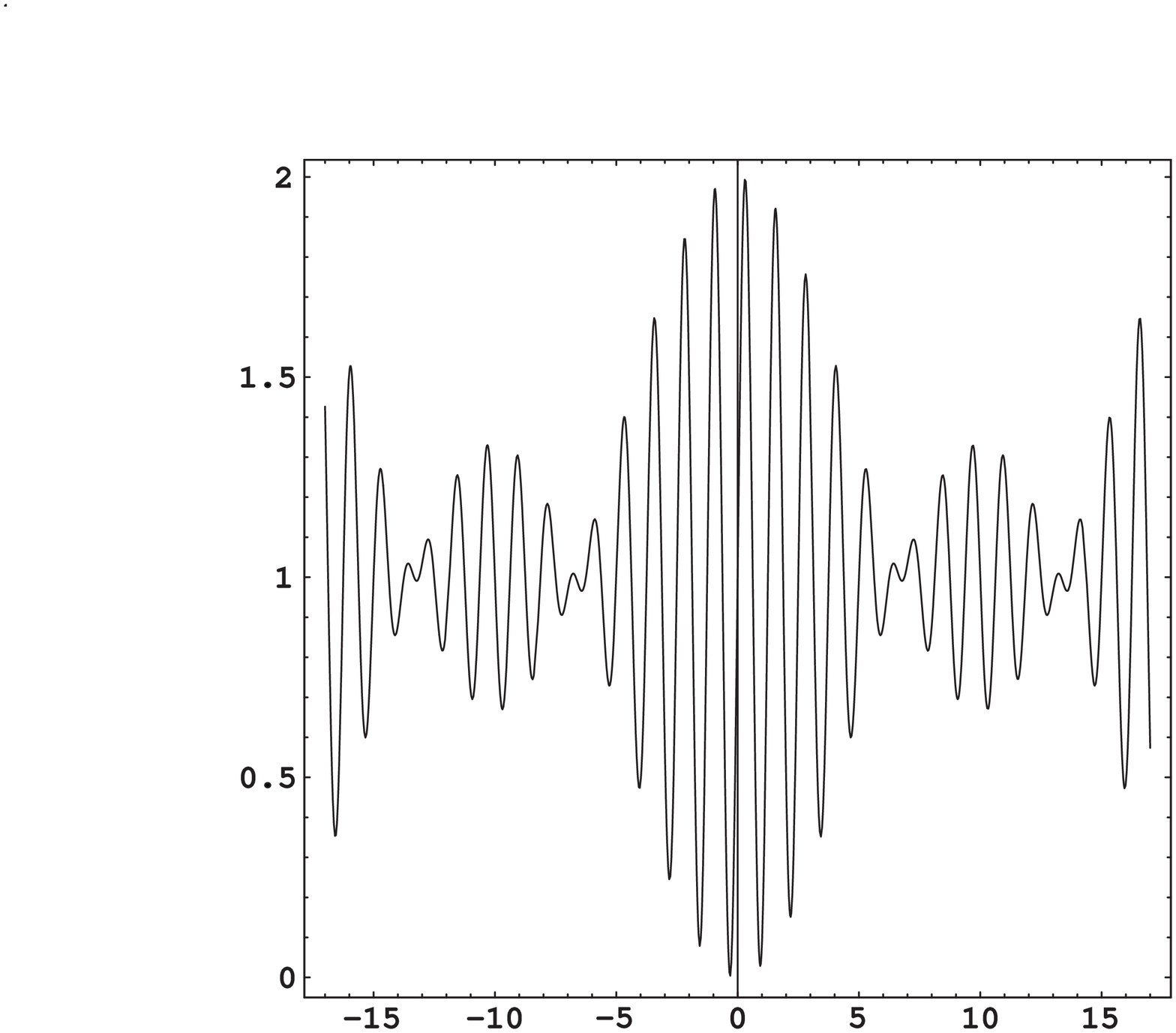,width=85mm,height=70mm}
\rotatebox{0}{\hskip 80pt  Optical pathlength difference,~$x$}
\end{minipage}
     \caption{Shown on the left are three   monochromatic
waves within the narrow finite bandwidth. Right plot shows interference
of the polychromatic light that was composed as a superposition of these three
waves. Note the envelope correction to the fringe visibility as a function of the OPD.
      \label{fig:jupiter_defl2}}
 \end{center}
\end{figure}

\begin{figure}[p] 
\begin{center} 
\rotatebox{90}{\hskip 50pt Fringe amplitude}
\hskip -15pt
\begin{minipage}[b]{.9\linewidth}
\vskip -25pt
\centering\epsfig{file=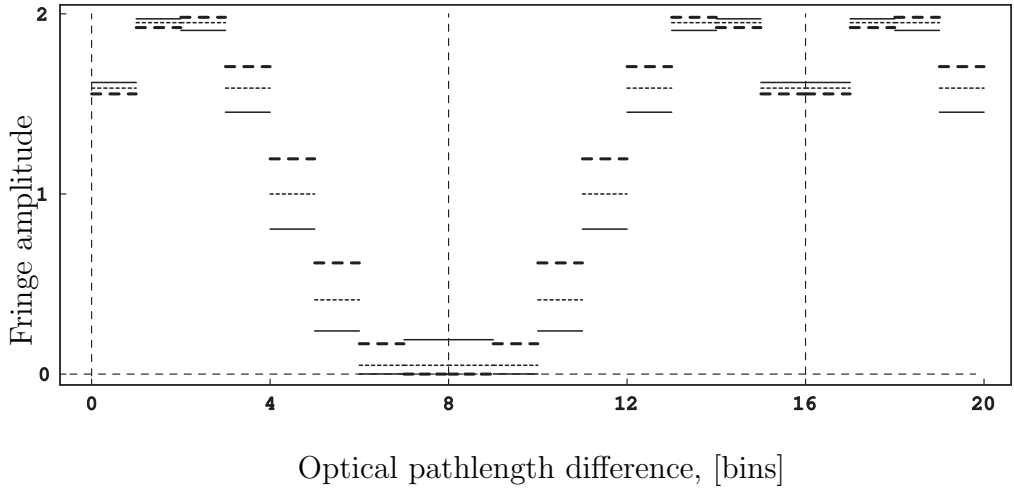,width=13cm}
\vskip -5pt \hskip -0pt
\rotatebox{0}{\hskip 17pt  Optical pathlength difference,~[bins] }
\end{minipage}
\vskip 35pt
\rotatebox{90}{\hskip 50pt Fringe amplitude}
\hskip -15pt
\begin{minipage}[b]{.9\linewidth}
\vskip -25pt
\centering\epsfig{file=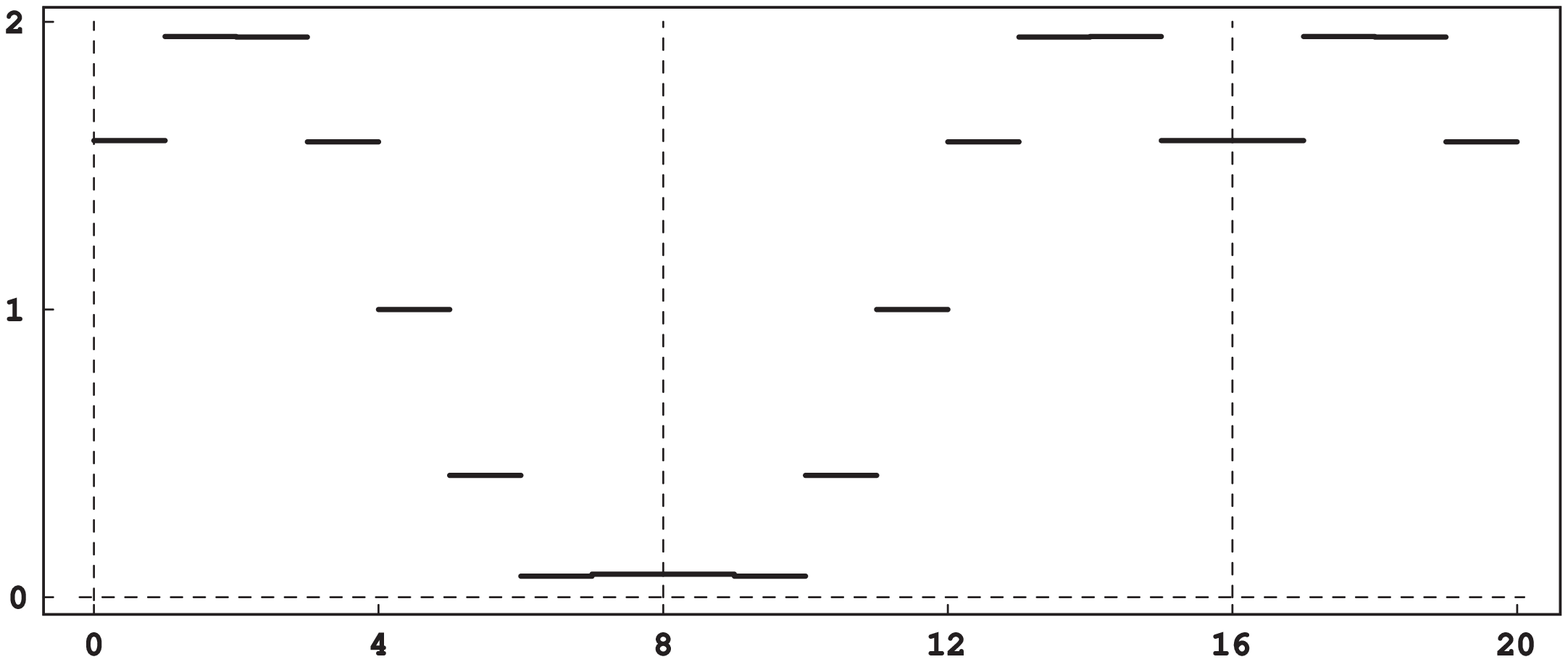,width=13cm}
\vskip -5pt \hskip -0pt
\rotatebox{0}{\hskip 17pt  Optical pathlength difference,~[bins] }
\end{minipage}
\caption{Upper plot shows typical behavior of three monochromatic sinusoidal fringes  
modulated in 8 equal steps over a wavelength (as shown in upper plot of Figure
\ref{fig:fringe_step}). Parameters for the waves are  given as $\phi= 
k_ix$, $\phi_0=0$, with $k_1=0.75k_0$ (thick dashed line), $k_2=0.8k_0$ (thin dashed
line), $k_3=0.85k_0$  (thin solid line), thus the width of the spectral channel is
$\Delta k=0.1k_0$. Lower plot shows polychromatic fringe composed from these three
waves.
\label{fig:fringe_step_p}}
\end{center}
\end{figure}

What makes the polychromatic case much more difficult to study is that the
different constituents of the finite   bandwidth  are not responding to the phase
modulation in the same way due to a short  coherence length, as shown in 
Figures \ref{fig:jupiter_defl2} and  \ref{fig:fringe_step_p}. In particular, 
Figure \ref{fig:jupiter_defl2} shows a typical interference pattern of the
polychromatic light that was composed as a superposition of a number of
monochromatic constituents. Also, in the upper plot of Figure \ref{fig:fringe_step_p}
we show behavior of three monochromatic sinusoidal fringes  plotted as functions of
OPD. The OPD  is modulated   in equal steps over a wavelength as shown in the upper
plot of Figure \ref{fig:fringe_step}. Note that largest wavenumber (plotted by thin
dashed line) produces the faster changing  fringe amplitude.  Moreover, the
polychromatic fringe composed from these three waves does not exactly repeats the
behavior of either of its  constituents. As a result, one would need to minimize the
bandwidth $\Delta k $ in order to be able to describe the polychromatic
phenomena. 

The integration over time may be performed in a general form  and corresponding
expression for the photon count,
$N_{\ell i}$, is given  as follows:
\begin{eqnarray}
N_{\ell i} &=&   {\cal I}_{0 \ell} 
\bigg(1 \,+  V_{0\ell}\sin\phi(k_\ell)\, {\sf Re}\big\{{\tilde{\cal P}}_{\ell i}\big\} +
V_{0\ell}\cos\phi(k_\ell)\, {\sf Im}\big\{{\tilde{\cal P}}_{\ell i}\big\}\bigg),
\label{eq:dtintf}
\end{eqnarray}
\noindent 
with quantities ${\sf Re}\big\{{\tilde{\cal P}}_{\ell i}\big\}$ and 
${\sf  Im}\big\{{\tilde{\cal P}}_{\ell i}\big\}$ given as below:
\begin{eqnarray}
{\sf Re}\big\{{\tilde{\cal P}}_{\ell i}\big\}&=&
\frac{1}{\Delta\tau_i}\int_{t^{-}_i}^{t^{+}_i} \!\!
{\cal E}_{\ell}\big[\Delta k_\ell, \phi(k_\ell), x(t)\big]
\cos\big[k_\ell x(t)+
\Omega_{\ell}[\Delta k_\ell, \phi(k_\ell), x(t)]\,\big]dt,\nonumber\\[0pt] 
{\sf Im}\big\{{\tilde{\cal P}}_{\ell i}\big\}&=&
\frac{1}{\Delta\tau_i}\int_{t^{-}_i}^{t^{+}_i} \!\!
{\cal E}_{\ell}\big[\Delta k_\ell, \phi(k_\ell), x(t)\big]
\sin\big[k_\ell x(t)+\Omega_{\ell}[\Delta k_\ell, \phi(k_\ell), x(t)]\,\big]dt.
\hskip 20pt 
\label{eq:n0jnewsec}
\end{eqnarray} 
For convenience of further analysis, we combined these two real-valued matrices 
into one complex matrix ${\tilde{\cal P}}_{\ell i}$  
\begin{eqnarray}
{\tilde{\cal P}}_{\ell i}&=&{\sf Re}\big\{{\tilde{\cal P}}_{\ell i}\big\}+j\,\,
{\sf Im}\big\{{\tilde{\cal P}}_{\ell i}\big\}.
\label{eq:Rli}
\end{eqnarray}
Furthermore, with  definitions for  ${\sf Re}\big\{{\tilde{\cal P}}_{\ell i}\big\}$ and
${\sf Im}\big\{{\tilde{\cal P}}_{\ell i}\big\}$,  Eq. (\ref{eq:n0jnewsec}),
this complex matrix may be presented as given below
\begin{eqnarray}
{\tilde{\cal P}}_{\ell i}&=&
\frac{1}{\Delta\tau_i}\int_{t^{-}_i}^{t^{+}_i} \!\!
{\cal E}_{\ell}\big[\Delta k_\ell, \phi(k_\ell), x(t)\big]
e^{j\, \big( k_\ell x(t)+
\Omega_{\ell}[\Delta k_\ell, \phi(k_\ell), x(t)]\,\big) }dt,
\label{eq:pmat}
\end{eqnarray}
which, with the help of Eq. (\ref{eq:wjs11}), is conveniently transforms as
follows
\begin{eqnarray}
{\tilde{\cal P}}_{\ell i}&=&
\frac{1}{\Delta\tau_i}\int_{t^{-}_i}^{t^{+}_i} \!\!
e^{ j\, k_\ell x(t) } \,\,{\tilde W_\ell}
\big[\Delta k_\ell, \phi(k_\ell), x(t)\big]dt,
\label{eq:Rli0lz}
\end{eqnarray}
where the  complex envelope function ${\tilde W_\ell}$ given by Eqs.
(\ref{eq:wj})-(\ref{eq:wj?}). 

Eq. (\ref{eq:Rli0lz}) may further be transformed to
establish its true dependency of the integrand on time and wavenumber. To do this, we
substitute expression for complex envelope function, Eq. (\ref{eq:wj?}), 
directly into Eq. (\ref{eq:Rli0lz}) and obtain the following form for matrix
${\tilde{\cal P}}_{\ell i}$
with explicit dependency  of the integrand:
\begin{eqnarray}
{\tilde{\cal P}}_{\ell i}&=&
 \frac{1}{{\cal I}_{0 \ell}V_{0\ell} \Delta k_\ell \Delta\tau_i}
\int_{t^{-}_i}^{t^{+}_i} \!\!
\int_{k^{-}_\ell}^{k^{+}_\ell} \!\!  
{\cal F}(k){\cal I}_0(k)V(k)\,
e^{j\,\big( k\,x(t)+\phi(k)-\phi(k_\ell)\big)} dt  dk.
\label{eq:wjc*}
\end{eqnarray}


Finally, we have defined everything that is needed to study Eq. (\ref{eq:dtintf}),
for the polychromatic fringe, which   may equivalently be presented in a matrix form as
below 
\begin{eqnarray}
\begin{array}({c})
N_{\ell 1} \\[0pt]
... \\[0pt]
N_{\ell N} 
\end{array}
&=& 
\begin{array}({ccc}) 
1; &~{\sf Im}\big\{{\tilde{\cal P}}_{\ell 1}\big\}; & 
~{\sf Re}\big\{{\tilde{\cal P}}_{\ell 1}\big\}\\[0pt]
... & ...& ...\\[0pt]
1; &~{\sf Im}\big\{{\tilde{\cal P}}_{\ell N}\big\}; &
~{\sf Re}\big\{{\tilde{\cal P}}_{\ell N}\big\} 
\end{array} 
\begin{array}({c})
{\cal I}_{0\ell}  \\[0pt]
{\cal I}_{0\ell} V_{0\ell}\cos  \phi(k_\ell) \\[0pt]
{\cal I}_{0\ell} V_{0\ell}\sin  \phi(k_\ell)  
\end{array},
\label{eq:wj12b*z}
\end{eqnarray}
where  the complex matrix ${\tilde{\cal P}}_{\ell i}$ is given by Eq.
(\ref{eq:wjc*}).

The obtained result Eq. (\ref{eq:dtintf}), (\ref{eq:wjc*}) (or, equivalently   Eq.
(\ref{eq:wj12b*z})) constitutes the general form of expression for the polychromatic
fringe. We will use this result to finalize the development of the general from of the
observational model for polychromatic case with  arbitrary phase modulation.

Ideally, one would need to determine  not only three quantities  
${\cal I}_{0\ell} V_{0\ell}\cos  \phi(k_\ell)$, ${\cal I}_{0\ell} V_{0\ell}\sin 
(k_\ell),$ and  ${\cal I}_{0\ell}$, but the full functional dependency of the original
quantities. However, the finite width of the observational band-width
$\Delta k_\ell$  complicates the estimation process by bringing the non-linearity in
the observational equation via the envelop function $W$. Note that if one  neglects
the size of the bandwidth
$\Delta k_\ell$ with respect to the mean wavenumber $  k_\ell$ or  
$\Delta k_\ell/{k_\ell}\rightarrow0$ (the envelop function becomes unity
$W\rightarrow 1$), one recovers the full simplicity of the monochromatic
case represented by Eq. (\ref{eq:N}).\cite{Turyshev2000}  

\section{General Solution for Polychromatic Phasors With Noisy Data}
\label{sec:pol_g} 

Currently in use, there are two fringe estimators, one for visibility (the
unbiased estimator is $V^2$), and one for the phase (the unbiased estimator is
the complex phasor). The $V^2$ estimator is already worked out in much detail
(i.e. Refs. \citeonline{markscott},\citeonline{colavita2}-\citeonline{goodman})  if
the  complex phasor estimator is completed. So the development of the complex phasor
was the main purpose for the presented work. As it is known, the complex fringe
visibility can be represented by a phasor; if the fringe is stable, we can add the
phasors vectorially over multiple samples. This co-adding can provide an improved
signal-to-noise ratio. To co-add the fringe phasors requires a phase reference, for
instance the white light  phase.\cite{shao,shao1} 

In this  Section  we will develop optimally-weighted solution that accounts for a
number of noise sources and will be applicable for a general case of
delay modulation.  

\subsection{Fringe Equation for Noisy Data}

For the purposes of clarity we will omit spectral index
$\ell$. All the obtained results are valid for any channel and thus could be easily
reconstructed, if needed.

In the case of noisy data,   observations of photon-counts $N_i$ are actually done
with errors and, in reality, we observe   $N_i={\bar N}_i+\epsilon_i$, where
${\bar N}_i$ is the value of photon counts  at the $i$-th temporal bin in the absence
of noise and $\epsilon_i$ is a random variable, representing the noise contribution to
the measurements. We assume  that
$\epsilon_i$ are random variables that are primarily due to   gaussian  statistics.
(This approach may be extended to incorporate other sources of noise.\cite{mct} The
corresponding results will be reported elsewhere.) that are distributed around zero
and following relations are valid
\begin{equation}
N_i={\bar N}_i+\epsilon_i, \qquad\qquad E(\epsilon_i)=0, 
\qquad\qquad E(\epsilon_i^2)=\sigma_i^2.
\end{equation}
In the general case one must account  not only for the Gaussian
statistics of read-out process, but also for the  Poisson statistic that governs  
photo-emission (and photo-counting) process. Thus, a correct approach would be to
assume that 
$\epsilon_i$ is a sum of two terms $\epsilon_i=\mu\,\epsilon^G_i+
(1-\mu)\,\epsilon^P_i$, where $\epsilon^G_i$ is  Gaussian  and  $\epsilon^P_i$ is
Poissonian variables and $\mu$ is a number between 0 and 1.  Expected
complication arises from the fact that  standard deviation computed for the
photon-counting Poissonian bias  is actually proportional to the signal
${(\sigma^P_i)}^2\propto {\bar{\cal I}}_i$ (see discussion in
Ref. \citeonline{colavita2}). This issue is out of scope of the present paper and we
will address this issue at a later time.

We also assume that $N_i$ are independent, therefore, we may form a diagonal
covariance matrix for the quantities  $N_i$ (or equivalently for $\epsilon_i$) with
dispersions $\sigma_i^2$ on the diagonal: 
\begin{equation}
C_y= 
\begin{array}({cccc})
\sigma_1^2;& 0;& ... &0\\
0; & \sigma_2^2; &... & 0\\
...& ...& ...& 0\\
0; & 0;& ... & \sigma_N^2
\end{array}, \qquad
G_y=C^{-1}_y= 
\begin{array}({cccc})
\sigma_1^{-2};& 0;& ... &0\\
0; & \sigma_2^{-2}; &... & 0\\
...& ...& ...& 0\\
0; & 0;& ... & \sigma_N^{-2}
\end{array}, 
\label{eq:cov_p}
\end{equation}
\noindent where $G_y$ is the  matrix of weights.
Therefore, in the case when noise is present in the data,  
equation (\ref{eq:dtintf}) has  following 
matrix form as below 
\begin{eqnarray}
{\bar N}_i + \epsilon_i 
&=& \Big(
1; ~~{\sf Im}\big\{{\tilde{\cal P}}_{\ell i} \big\};  
~~{\sf Re}\big\{{\tilde{\cal P}}_{\ell i} \big\}\Big) 
\begin{array}({c})
{\cal I}_{0\ell}  \\[0pt]
{\cal I}_{0\ell} V_{0\ell}\cos  \phi(k_\ell) \\[0pt]
{\cal I}_{0\ell} V_{0\ell}\sin  \phi(k_\ell)  
\end{array},
\label{eq:wj12b*q}
\end{eqnarray}
\normalsize
or,  equivalently,
\begin{equation}
{\bar N}_i+\epsilon_i=  A_{i\alpha}X^\alpha,
\label{eq:bs_gp}
\end{equation}

\noindent  with indexes $i$ and $\alpha$  running as $i\in\{1,..., N\}$ and
$\alpha\in\{1,2,3\}$. Vector $X^\alpha \equiv \Big({\cal I}_{0\ell};\,
{\cal I}_{0\ell} V_{0\ell}\cos  \phi(k_\ell);\,
{\cal I}_{0\ell} V_{0\ell}\sin  \phi(k_\ell)\Big)^T$ is the vector to be determined and
matrix  
${\bf A}^T=A_{i\alpha}\equiv \big(1; {\sf Im}\big\{{\tilde{\cal P}}_{\ell i} \big\};  
{\sf Re}\big\{{\tilde{\cal P}}_{\ell i} \big\}\big) $ is the $3\times N$  rotational
matrix   in the phase space.
A maximum likelihood solution to the system
of equations (\ref{eq:bs_gp}) may be given   by the following system of
equations 
\begin{equation}
X^\alpha= \sum_i^N {{\bf A}^\dagger_\diamond}^{i\alpha}{\bar N}_i, \qquad \qquad
{\rm where } \qquad \qquad
{{\bf A}^\dagger_\diamond}= ({\bf A}^TG_y{\bf A})^{-1}{\bf A}^TG_y,
\label{eq:formofsol_g_p}
\end{equation}
\noindent 
with  ${{\bf A}^\dagger_\diamond}$ being an optimally-weighted pseudo-inverse
matrix. Note that by choosing different gain matrix\cite{markscott} instead of
optimally weighted least-squared matrix Eq. (\ref{eq:cov_p}), one may obtain solution
with different, specifically designed properties. Nevertheless, our solution has 
enough embedded generality as it allows for arbitrary properties of noise
contribution, which will be further explored below. 

\subsection{Optimally-Weighted Pseudo-Inverse Matrix}
\label{sec:mon_pseudo_p} 

In this Section we will find solution for the pseudo-inverse matrix 
${{\bf A}^\dagger_\diamond}$ that was introduced by Eq.(\ref{eq:formofsol_g_p}).  
To construct this matrix we will use the weights matrix $G_y$ given by 
Eq.(\ref{eq:cov_p}) and the rotation  matrix  ${\bf A}_\diamond$  given by 
Eq.(\ref{eq:Rli0lz}) as: 
\begin{eqnarray}
 {\bf A}_{i} &=& \Big( 1;~\,{\sf Im}\big\{{\tilde{\cal P}}_{ i}\big\}; 
~{\sf Re}\big\{{\tilde{\cal P}}_{i}\big\}\Big) \equiv
 \Big( 1; ~s_i;  ~c_i\Big),
\label{eq:Rmat_g0p}
\end{eqnarray}
where we denoted $s_i={\sf Im}\big\{{\tilde{\cal P}}_{ i}\big\},  
c_i ={\sf Re}\big\{{\tilde{\cal P}}_{  i}\big\}$.
 
Let us   construct  matrix $({\bf A}^TG_y{\bf A})$  first. Calculation of $({\bf A}^TG_y{\bf A})$ is straightforward  even for the most general case of arbitrary
number of temporal bins ($N\ge3$) and with arbitrary integration intervals ($\Delta
\tau_i\not=\Delta \tau_j$  for  $i\not=j$). Thus, after some algebra we find the
following structure:
\begin{equation}
({\bf A}^TG_y)=
\begin{array}({c})
\frac{1}{\sigma_1^2}; ~\frac{1}{\sigma_2^2};
~... ~\frac{1}{\sigma_N^2}  \\[0pt]
\frac{s_1}{\sigma_1^2}; ~\frac{s_2}{\sigma_2^2};
~...  ~\frac{s_N}{\sigma_N^2}  \\[0pt]
\frac{c_1}{\sigma_1^2}; ~\frac{c_2}{\sigma_2^2}; 
~... ~\frac{c_N}{\sigma_N^2}  
\end{array}, ~~  
({\bf A}^TG_y{\bf A}) ~= ~
\begin{array}({lll})
\overset{N}{\underset{i}\sum}\frac{1}{\sigma_i^2}
&\overset{N}{\underset{i}\sum} \frac{s_i}{\sigma_i^2}
&\overset{N}{\underset{i}\sum}\frac{c_i}{\sigma_i^2}\\[0pt]
\overset{N}{\underset{i}\sum}\frac{s_i}{\sigma_i^2}&
\overset{N}{\underset{i}\sum} \frac{s_i^2}{\sigma_i^2}&
\overset{N}{\underset{i}\sum}\frac{s_ic_i}{\sigma_i^2} \\[0pt]
\overset{N}{\underset{i}\sum}\frac{c_i}{\sigma_i^2} &
\overset{N}{\underset{i}\sum}\frac{s_ic_i}{\sigma_i^2} &
\overset{N}{\underset{i}\sum}\frac{c_i^2}{\sigma_i^2}\\
\end{array}.
\end{equation}

\noindent By inverting the obtained result one constructs  the covariance matrix
$\Lambda$ of the following structure: 
\begin{eqnarray}
&&\hskip 20pt 
{\bf \Lambda}~=~({\bf A}^TG_y{\bf A})^{-1}~= \nonumber\\[10pt] 
&&\hskip -30pt 
=~\frac{1}{\Delta_\diamond}
\begin{array}({lll})
~~\,\frac{1}{2}\overset{N}{\underset{ij}\sum}
\frac{(s_ic_j-c_is_j)^2}{\sigma_i^2\sigma_j^2}
&~~\frac{1}{2}\overset{N}{\underset{ij}\sum}\frac{(c_i-c_j)(s_ic_j-c_is_j)}
{\sigma_i^2\sigma_j^2}
&-\frac{1}{2}\overset{N}{\underset{ij}\sum}\frac{(s_i-s_j)(s_ic_j-c_is_j)}
{\sigma_i^2\sigma_j^2}\\[10pt]
~~\frac{1}{2}\overset{N}{\underset{ij}\sum}\frac{(c_i-c_j)(s_ic_j-c_is_j)}
{\sigma_i^2\sigma_j^2}
&~~\frac{1}{2}
\overset{N}{\underset{ij}\sum}\frac{(c_i-c_j)^2}{\sigma_i^2\sigma_j^2}
&-\frac{1}{2}\overset{N}{\underset{ij}\sum}\frac{(s_i-s_j)(c_i-c_j)}
{\sigma_i^2\sigma_j^2}\\[0pt]
-\frac{1}{2}\overset{N}{\underset{ij}\sum}\frac{(s_i-s_j)(s_ic_j-c_is_j)}
{\sigma_i^2\sigma_j^2} 
&-\frac{1}{2}\overset{N}{\underset{ij}\sum}\frac{(s_i-s_j)(c_i-c_j)}
{\sigma_i^2\sigma_j^2}
&~~\,\frac{1}{2}\overset{N}{\underset{ij}\sum}
\frac{(s_i-s_j)^2}{\sigma_i^2\sigma_j^2}\nonumber
\end{array},
\end{eqnarray}
\normalsize

\noindent where  determinant of the matrix $ ({\bf A}^TG_y{\bf A})$,
$\Delta_\diamond=\det||({\bf A}^TG_y{\bf A})||$, is given as 
\vskip -15pt
\begin{eqnarray}
\Delta_\diamond&=&\frac{1}{2}\overset{N}{\underset{ijk}\sum}~
\frac{(s_ic_j- s_jc_i)}{\sigma_i^2\sigma_j^2\sigma_k^2}
\Big[(s_ic_j- s_jc_i)+(s_jc_k- s_kc_j)+(s_kc_i- s_ic_k)\Big],
\label{eq:det_g_p} 
\end{eqnarray}
with a triple summation  for all the indexes denoting the temporal bins and
 running from  1 to $N$, namely $\forall
~~\{i,j,k\}\in [ 1, ..., N]$.

These intermediate results allow us to write the 
solution for the $(N\times 3)$ optimally-weighted 
pseudo-inverse matrix 
${{\bf A}^\dagger_\diamond}={A_\diamond^\dagger}^\alpha_k$ 
in the following compact  form: 
\begin{equation}
{{\bf A}^\dagger_\diamond}= ({\bf A}^TG_y{\bf A})^{-1}{\bf A}^TG_y
= \frac{1}{{\cal D}^\diamond}
\begin{array}({c})
{\cal A}^\diamond_k\\[-2pt]
{\cal B}^\diamond_k\\[-2pt]
{\cal C}^\diamond_k
\end{array},
\label{eq:r_dag_g_p}
\end{equation}

\noindent where   coefficients ${\cal A}^\diamond_{k}, {\cal B}^\diamond_k, 
{\cal C}^\diamond_k$  and ${\cal D}^\diamond$  depend  on   duration of
each temporal bin,  mean wavenumber  and variances
for the data taken in each bin, and are  given by
\begin{eqnarray} 
{\cal A}_{k}^\diamond &=&
 { \overset{N}{\underset{ij}\sum}~\frac{1}{\sigma_i^2\sigma_j^2\sigma_k^2}
\big(s_ic_j- s_jc_i\big) 
\big[(s_ic_j- s_jc_i)+(s_jc_k- s_kc_j)+(s_kc_i- s_ic_k)\big]},
\nonumber\\[6pt]
{\cal B}_{k}^\diamond&=&
 {  \overset{N}{\underset{ij}\sum}~\frac{1}{\sigma_i^2\sigma_j^2\sigma_k^2}
\big( c_i-c_j\big)~ 
\big[(s_ic_j- s_jc_i)+(s_jc_k- s_kc_j)+(s_kc_i- s_ic_k)\big]},
\nonumber\\[6pt]
{\cal C}_{k}^\diamond &=& -~ 
 { \overset{N}{\underset{ij}\sum}~\frac{1}{\sigma_i^2\sigma_j^2\sigma_k^2}
\big( s_i-s_j\big)~ 
\big[(s_ic_j- s_jc_i)+(s_jc_k- s_kc_j)+(s_kc_i- s_ic_k)\big]},
\nonumber\\[6pt]
{\cal D}^\diamond&=&\overset{N}{\underset{k}\sum}~{\cal A}^\diamond_{k}~=~
\frac{1}{3}\overset{N}{\underset{ijk}\sum}\frac{1}{\sigma_i^2\sigma_j^2\sigma_k^2}
\big[(s_ic_j- s_jc_i)+(s_jc_k- s_kc_j)+(s_kc_i- s_ic_k)\big]^2.
\hskip 20pt
\label{eq:phasorse2_g_p}
\end{eqnarray}
 
Definitions for the  quantities $s_i$ and $c_i$,  
$s_i={\sf Im}\big\{{\tilde{\cal P}}_{ i}\big\},   
c_i ={\sf Re}\big\{{\tilde{\cal P}}_{  i}\big\}$, allow to present
expressions (\ref{eq:phasorse2_g_p}) in a more convenient form.  First, remember 
that complex  matrix, ${\tilde{\cal P}}_{  i}$, as any
complex function, may be represented by its amplitude and its phase, namely 
\begin{eqnarray}
{\tilde{\cal P}}_{i}&=&{\sf Re}\big\{{\tilde{\cal P}}_{i}\big\}+j\,\,
{\sf Im}\big\{{\tilde{\cal P}}_{i}\big\} =
 p_{i} \,e^{j \pi_{i}},
\label{eq:wjs11app}
\end{eqnarray}
\noindent where $p_{i}$ and $\pi_{i}$ are the amplitude and  the
phase of this  complex matrix correspondingly and are given as follows:
\begin{eqnarray}
p_{i} =
 \sqrt{{\sf Re}^2\big\{{\tilde{\cal P}}_{i}\big\}+ 
{\sf Im}^2\big\{{\tilde{\cal P}}_{i}\big\}}, 
\qquad 
\pi_i= {\sf ArcTan}\Big\{
\frac{{\sf Im}\big\{{\tilde{\cal P}}_{i}\big\}}
{{\sf Re}\big\{{\tilde{\cal P}}_{i}\big\}}\Big\},
\label{eq:wjlls_app}
\end{eqnarray}
with   complex matrix ${\tilde{\cal P}}_{\ell i}$ is given by Eq. (\ref{eq:Rli0lz}).
These quantities allow presentation of Eqs. (\ref{eq:phasorse2_g_p}) in
the following form:
\begin{eqnarray} 
{\cal A}_{k}^\diamond &=&
 \overset{N}{\underset{ij}\sum}~
\frac{p_ip_j\sin[\pi_i-\pi_j]}{\sigma_i^2\sigma_j^2\sigma_k^2} 
\nonumber\times\\
&&\hskip 25pt\times
\Big(p_ip_j\sin[\pi_i- \pi_j]+p_jp_k\sin[\pi_j- \pi_k]+p_kp_i\,\sin[\pi_k-
\pi_i]\Big),
\nonumber\\[6pt]
{\cal B}_{k}^\diamond&=&
  \overset{N}{\underset{ij}\sum}~
\frac{\big( p_i\cos \pi_i-p_j\cos \pi_j\big)}{\sigma_i^2\sigma_j^2\sigma_k^2} 
\nonumber\times\\
&&\hskip 25pt\times
\Big(p_ip_j\sin[\pi_i- \pi_j]+p_jp_k\sin[\pi_j- \pi_k]+p_kp_i\sin[\pi_k-
\pi_i]\Big),
\nonumber\\[6pt]
{\cal C}_{k}^\diamond &=& -~ 
 \overset{N}{\underset{ij}\sum}~
\frac{\big( p_i\sin \pi_i-p_j\sin \pi_j\big)}{\sigma_i^2\sigma_j^2\sigma_k^2} 
\nonumber\times\\
&&\hskip 25pt\times
\Big(p_ip_j\sin[\pi_i- \pi_j]+p_jp_k\,\sin[\pi_j- \pi_k]+p_kp_i\,\sin[\pi_k-
\pi_i]\Big),
\nonumber\\[6pt]
{\cal D}^\diamond&=&\overset{N}{\underset{k}\sum}~{\cal A}^\diamond_{k}=
\nonumber\\
&=& 
\frac{1}{3}\overset{N}{\underset{ijk}\sum}\frac{1}{\sigma_i^2\sigma_j^2\sigma_k^2}
\Big(p_ip_j\sin[\pi_i- \pi_j]+p_jp_k\sin[\pi_j- \pi_k]+p_kp_i\sin[\pi_k-
\pi_i]\Big)^2.
\hskip 10pt
\label{eq:phasorse2_g}
\end{eqnarray}
 
At this point we have all the expressions necessary to present the
optimally-weighted solution for the polychromatic phasors.  

\subsection{Photon Noise-Optimized Solution for Polychromatic Phasors}
\label{sec:mon_phasors_g_p} 

An optimally-weighted solution for  the quantities $X^\alpha$ may  be
obtained directly now from Eq.(\ref{eq:formofsol_g_p}) with the help of expressions  
(\ref{eq:r_dag_g_p})-(\ref{eq:phasorse2_g_p})  in the following compact form: 
\small
\begin{eqnarray} 
{\cal I}^\diamond_0 &=&\frac{1}{{\cal D}^\diamond}\overset{N}{\underset{k}\sum}~
{\bar N}_k~{\cal A}^\diamond_{k},
\nonumber\\[6pt]
{\cal I}^\diamond_0~ V_0^\diamond\cos\bar{\phi}^\diamond&=&\frac{1}{{\cal
D}^\diamond}\overset{N}{\underset{k}\sum}~ {\bar N}_k~
 {\cal B}^\diamond_{k},
\nonumber\\[6pt]
{\cal I}^\diamond_0~ V_0^\diamond\sin\bar{\phi}^\diamond&=& \frac{1}{{\cal
D}^\diamond}
\overset{N}{\underset{k}\sum}~{\bar N}_k~{\cal C}^\diamond_{k}.
\label{eq:phasorse1_g_p}
\end{eqnarray}
\normalsize
\noindent with coefficients of ${\cal A}^\diamond_k, {\cal B}^\diamond_k, 
{\cal C}^\diamond_k$   and ${\cal D}^\diamond$ are given by
Eqs.(\ref{eq:phasorse2_g_p}) and (\ref{eq:phasorse2_g}). 

The obtained solution for the polychromatic visibility phasors given by Eq.
(\ref{eq:phasorse1_g_p}) is given in the form of a linear combination of weighted
photon counts recorded during a particular integration period. This   form turned
out to be very helpful when analyzing contributions of CCD pixels that are
systematically biased. The obtained result may be used to de-weight 'bad' pixels (in a
statistical sense) and, thus, to reduce the problem of biases while estimating fringe
parameters.

This form allows to express an optimally-weighted solution for visibility, phase
and the constant intensity terms in a familiar compact form:
\begin{eqnarray} 
{V^\diamond_{0}}^2&=&\frac{\big(\overset{N}
{\underset{k}\sum}~{\bar N}_k\,{\cal B}^\diamond_{k}\big)^2+
\big(\overset{N}{\underset{k}\sum}~{\bar N}_k\,{\cal C}^\diamond_{k}\big)^2}
{\big(\overset{N}{\underset{k}\sum}~{\bar N}_k\,{\cal
A}^\diamond_{k}\big)^2},\nonumber\\[6pt]
\bar{\phi}^\diamond
&=&\text{ArcTan}\Big[\,\frac{\overset{N}{\underset{k}\sum}~{\bar N}_k~ 
{\cal C}^\diamond_{k}}{\overset{N}{\underset{k}\sum}~{\bar N}_k~ 
{\cal B}^\diamond_{k}}\,\Big], \qquad ~~  {\cal I}^\diamond_0
=\frac{\overset{N}{\underset{k}\sum}~{\bar N}_k~{\cal A}^\diamond_{k}}
{\overset{N}{\underset{k}\sum}~{\cal A}^\diamond_{k}}. 
\label{eq:phasorse3_g_p}
\end{eqnarray}

The form of the obtained solution is  simple to understand and it is straightforward 
to implement in the software codes. All the information necessary to calculate the $3N$
coefficients of ${\cal A}^\diamond_k, {\cal B}^\diamond_k, {\cal C}^\diamond_k$ 
and ${\cal D}^\diamond$ is presumed to be known  before the  experiment.  Thus,
for the case when $N=8$ one would have to calculate  only 24 numbers from Eq.
(\ref{eq:phasorse2_g_p}). These numbers correspond to  8 numbers of ${\cal
A}^\diamond_k$, 8 numbers of
${\cal B}^\diamond_k$ and  8 numbers of ${\cal C}^\diamond_k$. 
The experimental data is used as input to Eqs.~(\ref{eq:phasorse1_g_p}) (or directly
Eqs.~(\ref{eq:phasorse3_g_p})) to produce the best estimates of the actual values
for visibility, phase and the constant intensity term.  
This approach is currently being utilized and  corresponding results will be reported
elsewhere.

\section{Filtered Polychromatic Light: Spectral Channels with Narrow Bands}
\label{sec:filter}

It is thought now that  SIM will be operating at 80 spectral channels for the
science interferometer and at 4 channels for both guide interferometers. The
data will be read at a millisecond time rate. SIM will be dispersing light just before
the interfering radiation reaches the detector.  To take this fact into account,  we
define a filter that allows  to limit the total bandwidth  of the incoming radiation.
We will formally denote a filter with such a properties as
follows:
\begin{equation}
 {\cal F}_\ell(k)={\cal F}_\ell(k-k_\ell; \Delta k_\ell).
\label{eq:filter}
\end{equation}
We assumed that  this filter operates within the ${\ell}$-th spectral channel by
allowing to pass through only such a radiation that is composed from the frequencies
corresponding to the mean (or central) wavenumber $k_\ell$ within the bandwidth of
$\Delta k_\ell$. Thus, filter ${\cal F}_\ell(k)$   enables the  instrument to ``see'' 
light only in a certain interval $\Delta k_\ell$ around the mean number
$k_\ell$. 

In this Section we will develop a model that employes such an approach for the
case of a rectangular bandpass filter. Due to its analytical simplicity, 
the rectangular bandpass filter is the most known construction in the Fourier
optics. This analysis will allow us to establish correspondence with the previously
obtained results both for  monochromatic and polychromatic light.

\subsection{Rectangular Bandpass Filter}
\label{sec:rect}

To take advantage of the results derived in the previous section, we must first
decide on the properties of the bandpass filter. This decision in return will affect
the    properties of the envelop function.  Below we shall develop a model for a
special case of the bandpass filter -- a rectangular bandpass filter denoted here as
${\cal F}_\ell$, which is done analytically  in the following form
\begin{equation}
 {\cal F}(k)=\sum_{\ell=1}^L {\cal F}_\ell(k), \qquad 
{\rm where } \qquad{\cal F}_\ell(k)=
\begin{cases} 
~~ {{\cal F}_{0\ell}}\,=\,{\rm const}, &
~~~~\text{ $k\in [k^-_{\ell}, k^+_{\ell}]$},\\[-10pt]
~~~~0,& ~~~~\text{ $k \not\in [k^-_{\ell}, k^+_{\ell}]$}.
\end{cases}
\label{eq:wj12b**z}
\end{equation}
\noindent 
We can also assume that the width of a spectral channel is small, so that both
intensity of incoming radiation, $ {\cal I}_{0}(k) $, and apparent visibility,
$V(k)$, do not change within the spectral channel (in particular, this leads to 
$\hat{V}_{0\ell}(k)\equiv1$ in Eqs. (\ref{eq:v0j}) and (\ref{eq:v0jno})).
Therefore, the following conditions are satisfied with a particular spectral
channel, $\ell$:
\begin{eqnarray}
{\cal I}_{0}(k)={\rm const}, \qquad V(k) &=& {\rm const}, 
\qquad {\cal F}_{0\ell}= {\rm const}, \\ 
\phi(k)-\phi(k_\ell)&=& d_{0\ell}(k-k_\ell)+
{\cal O}(\frac{\partial^2 \phi}{\partial k_\ell^2}), 
\end{eqnarray}
where $d_{0\ell}=\frac{\partial \phi}{\partial k_\ell}$ is the delay within the
$\ell$-th channel.

 One may perform integration of the fringe envelope
function  ${\tilde W_\ell}\big[x(t)\big]$ which is given by Eq. (\ref{eq:wj})
(or  use equation  (\ref{eq:wj8}) for the  unperturbed envelope function 
and then apply iterative procedure outlined in  Appendix \ref{sec:appa}.).
 To the second order in phase variation (i.e.
${\cal O}(\frac{\partial^2 \phi}{\partial k_\ell^2}) $), the resulted envelope
function has following properties:
\begin{eqnarray}
{\tilde W_\ell}\big[\Delta k_\ell, \phi(k_\ell), x_i\big]&=&
 \frac{1}{{\cal I}_{0 \ell}V_{0\ell} 
\Delta k_\ell}\int_{k^{-}_\ell}^{k^{+}_\ell} \!\!
{\cal F}(k){\cal I}_0(k)V(k)\,
e^{j\,\big((k-k_\ell)x(t)+\phi(k)-\phi(k_\ell)\big)} dk=\nonumber\\
&=&
 \frac{1}{\Delta k_\ell}
\int_{k^{-}_\ell}^{k^{+}_\ell} \!\!
e^{j\, (k-k_\ell)\big(x(t) \,+\,d_{0\ell}\big)} dk +
{\cal O}(\frac{\partial^2 \phi}{\partial k_\ell^2}). 
\label{eq:wjzzv}
\end{eqnarray}
   
Let us introduce a convenient variable,  $\kappa=k-k_\ell$, and  remember that $\Delta
k_\ell=k^+_{\ell}-k^-_{\ell}$ and 
$k_\ell=\frac{1}{2}(k^+_{\ell}+k^-_{\ell})$ are the width of the spectral channel and
the mean wavenumber. This allows us to integrate  expression (\ref{eq:wjzzv}) over the
wavenumber space 
\begin{eqnarray}
{\tilde W_\ell}\big[\Delta k_\ell, \phi(k_\ell), x_i\big]&=&
 \frac{1}{\Delta k_\ell}
\int_{-\frac{1}{2}\Delta k_\ell}^{+\frac{1}{2}\Delta k_\ell}  \!\!
e^{j\, \kappa\big(d_{0\ell}+x(t)\big) \!} d\kappa= 
\frac{\sin[\frac{1}{2}\Delta k_\ell(d_{0\ell}+x(t))]}
{\frac{1}{2}\Delta k_\ell(d_{0\ell}+x(t))}. 
\hskip20pt
\label{eq:wj?v}
\end{eqnarray}

We can now present Eq. (\ref{eq:Rli0lz}) for matrix ${\tilde{\cal P}}_{\ell i}$ as follows:
\begin{eqnarray}
{\tilde{\cal P}}_{\ell i}&=&
 \frac{1}{ \Delta\tau_i}
\int_{t^{-}_i}^{t^{+}_i} \!\! e^{ j\, k_\ell  x(t)  }
\,\frac{\sin[\frac{1}{2}\Delta k_\ell(d_{0\ell}+x(t))]}
{\frac{1}{2}\Delta k_\ell(d_{0\ell}+x(t))}\,dt .\label{eq:ugly03} 
\end{eqnarray}
At this moment, we introduce another convenient variable,  
$\tau=t-t_i$.  Analogously, $\Delta \tau_i=t^+_i-t^-_i$ and
$t_i=\frac{1}{2}(t^+_i+t^-_i)$   are the duration of the temporal integration
within the $i$-th bin and the mean time for this bin correspondingly.   This
result is used to transform  Eq. (\ref{eq:ugly03}) as below:
\begin{eqnarray}
{\tilde{\cal P}}_{\ell i}&=&\!\!
 {e^{ j\, k_\ell  x(t_i) }} ~
\delta{\tilde{\cal P}}_{\ell i}, 
\end{eqnarray}
with matrix coefficient $\delta{\tilde{\cal P}}_{\ell i} $ given by
\begin{eqnarray}
\delta{\tilde{\cal P}}_{\ell i}= \frac{1}{ \Delta\tau_i}
\int_{-\frac{1}{2}\Delta\tau_i}^{+\frac{1}{2}\Delta\tau_i} 
\!\!e^{ j\, k_\ell \big(x(t_i+\tau) -x(t_i)\big) }
\,\frac{\sin[\frac{1}{2}\Delta k_\ell\big(d_{0\ell}+x(t_i+\tau)\big)]}
{\frac{1}{2}\Delta k_\ell\big(d_{0\ell}+x(t_i+\tau)\big)}\,d\tau.
\label{eq:ugly04} 
\end{eqnarray}

  The obtained result explicitly depends on the functional form of the OPD
modulation, $x(t)$. To integrate this equation one first needs to make certain
assumptions on the temporal behavior of $x(t)$, which will be done in the following
Sections. At this moment we present Eq. (\ref{eq:dtintf}) in the following   form:
\begin{eqnarray}
N_{\ell i} &=&   {\cal I}_{0 \ell} 
\bigg(1 \,+   V_{0\ell}\sin\big[\phi(k_\ell)+k_\ell x(t_i)\big] 
\, {\sf Re}\big\{\delta{\tilde{\cal P}}_{\ell i}\big\} + \nonumber\\
&&\hskip 35pt +~
 V_{0\ell}\cos\big[\phi(k_\ell)+k_\ell x(t_i)\big] 
\, {\sf Im}\big\{\delta{\tilde{\cal P}}_{\ell i}\big\}\bigg),\hskip  10pt
\label{eq:dtint9s}
\end{eqnarray} 
where  the complex matrix $\delta{\tilde{\cal P}}_{\ell i}$ is given by  Eq.
(\ref{eq:ugly04}).  The importance of separating terms with $\delta{\tilde{\cal
P}}_{\ell i}$ is in the fact that one can establish clear correspondence with
monochromatic light, for which $\delta{\tilde{\cal P}}_{\ell i}=I_{\ell i}$, the
identity matrix.

In the next two subsections we will study two different special cases
of OPD modulation, namely the stepping  and ramping  modulations of the optical path
difference.

\subsection{Stepping Phase Modulation}

The stepping phase modulation realized when the  pathlength difference 
is changes as a set of discrete values corresponding to a number of steps
in the OPD space. mathematically this process represented as follows:  
\begin{equation}
x(t)=  \sum_{i=1}^{N=8} x(t_i), 
~~~~~~ \text{where} ~~~~~~
x(t_i)=
\begin{cases} 
x_i, & ~~~ t~\in [t^-_i,t_i^+],\\[-10pt]
 ~0, & ~~~ t~ \not\in [t^-_i,t_i^+],
\end{cases}
\label{eq:timesteps}
\end{equation}

\noindent with $t_i=\frac{1}{2}(t^+_i+t^-_i)$.
This procedure  defines the temporal bins
that will be used to modulate the interferometric pattern. 
 
Conditions (\ref{eq:timesteps})  allow for a significant 
simplification of  temporal integration in Eq. (\ref{eq:ugly04}).  It simply is
leading to a substitution $x(t)\rightarrow x_i$ in Eq. (\ref{eq:n0j}),
and matrix ${\tilde{\cal P}}_{\ell i}$  takes the following form
\begin{eqnarray}
{\tilde{\cal P}}_{\ell i}&=&
e^{ j\, k_\ell x(t_i) } \,\,
\frac{\sin[\frac{1}{2}\Delta k_\ell\big(d_{0\ell}+x_i\big)]}
{\frac{1}{2}\Delta k_\ell\big(d_{0\ell}+x_i\big)},
\label{eq:ugly04s} 
\end{eqnarray}
 As a result, to the second order in the phase variation
(i.e. $\phi(k_\ell)\approx\phi_\ell+{\cal O}(\frac{\partial^2
\phi}{\partial k_\ell^2}\Big|_{ k_\ell}) $), observational equation 
(\ref{eq:dtint9s}) is taking  form as  below:
\begin{eqnarray}
N_{\ell i} &=&  {\cal I}_{0 \ell} 
\bigg(1+V_{0\ell}\,
{\rm sinc}[\frac{1}{2}\Delta k_\ell\big(x_i+d_{0\ell}\big)]\,
\sin\big[\phi(k_\ell)+k_\ell x_i\big]\bigg).
\label{eq:n0j12q}
\end{eqnarray}

\noindent The obtained result   Eq. (\ref{eq:n0j12q}) clearly depends on the
particular form of the envelope function. As such, it has most of the parameters that
are necessary  for the phase estimation purposes in the case of wide bandwidth.  

For the most practical cases the value of the ${sinc}$
function will be close to $\rm sinc \sim1$. Indeed, let us analyze the argument of
this function, $\frac{1}{2}\Delta k_\ell\big(x_i+d_{0\ell})$.  Thus, one might
expect that within the spectral channel the phase will stay constant, hence 
$d_{0\ell}=\frac{\partial \phi}{\partial k_\ell} \approx 0$. Furthermore, for the
estimation purposes let us assume that all the step-sizes
$x_i$ are essentially $x_i=i\frac{\lambda_0}{N} $, where $N$ is the total number
of temporal integration bins, $i$ is the number of a particular temporal bin,
$i\in[1,...,N]$, and $\lambda_0$ is the reference wavelength chosen for the
OPD modulation   (or $\lambda_0 =\frac{2\pi}{k_0}$, where $k_0$ is the
wavenumber corresponding to chosen modulation wavelength). Also remember
that the width of a spectral channel is related to the total SIM bandwidth as 
$\Delta k_\ell =\frac{\Delta k_{\sf SIM} }{L}$, where $\Delta k_{\sf SIM}$ is
the total SIM  bandwidth and $L$ is the total number of spectral channels used for
the white light fringe detection. Therefore, one has
\begin{equation}
\frac{1}{2}\Delta k_\ell\big(x_i+d_{0\ell})\approx 
\frac{1}{2}\Delta k_\ell  x_i = 
\frac{\pi i}{LN} \frac{\Delta k_{\sf SIM} }{k_0}.
\label{eq:estim} 
\end{equation}
Assuming $\lambda_{\sf SIM}^-=450$ nm and $\lambda_{\sf SIM}^+ =900$ nm, and
$\lambda_0 =900$ nm,   thus yielding 
$\frac{\Delta k_{\sf SIM} }{k_0} =1.$ The maximal value for the expression 
(\ref{eq:estim}) is realized when $i=N$, thus  
\begin{equation}
\frac{\pi i}{LN} \frac{\Delta k_{\sf SIM} }{k_0}\le 
\frac{\pi }{L}.
\label{eq:estim1} 
\end{equation}
Currently, there are different numbers of spectral channels used to process data
from our testbeds. This number may be as large as $L=80$ and as small as $L=4$. 
Of coarse, when $L=80$, the ratio $\pi/L$ becomes $\pi/L=0.03927$ and, thus, 
${\sf sinc}[\frac{1}{2}\Delta k_\ell x_i ]|_{L=80}=0.99974$, and
similarly for $L=4$ the {sinc} function becomes 
${\sf sinc}[\frac{1}{2}\Delta k_\ell x_i ]|_{L=4}=0.90032$. We will
address the issue of phase estimation sensitivity to the width of a spectral
channel $\Delta k_\ell$ at a later time.  

This observation allows us to present the {\sf sinc} function as a series
with respect to the small parameter ${\Delta k_\ell}z$ (with $z$ being defined as
$z=x_i+d_{0\ell}$) as 
\begin{eqnarray}
{\rm sinc}[\frac{\Delta k_\ell z}{2}]&=&
\sum_{n=0}^\infty
\frac{(-1)^{n}}{(2n+1)!}\big[\frac{\Delta k_\ell z}{2}\big]^{2n}\nonumber\\
&=&
1-\frac{1}{3!}\big[\frac{\Delta k_\ell z}{2}\big]^2 +
\frac{1}{5!}\big[\frac{\Delta k_\ell z}{2}\big]^4+
{\cal O}( \frac{1}{7!}\big[\frac{\Delta k_\ell z}{2}\big]^6),
\label{eq:sin_exp}
\end{eqnarray}
\noindent one can present the expression Eq. (\ref{eq:n0j12q}) in the following
form:
\begin{eqnarray}
N_{\ell i} &=&  {\cal I}_{0 \ell} 
\Big[1+\nonumber\\
&+& V_{0\ell}\,
\Big(1-\frac{\Delta k_\ell^2(x_i+d_{0\ell})^2}{24} +
\frac{\Delta k_\ell^4(x_i+d_{0\ell})^4}{1920}\Big)\,
\sin\big[\phi(k_\ell)+k_\ell x_i\big]\Big].
\label{eq:step}
\end{eqnarray}


The obtained expression models the expected number of photons detected at the CCD for
the rectangular bandpass filter and stepping phase modulation. It extends the results
obtained for the monochromatic case on the finite size spectral bandwidth.  This fact
is indicated by the explicit dependency of the obtained result on the width of a
spectral channel $\Delta k_\ell$. (For the most of the interesting practical
applications, the size of the delay within a particular spectral channel is very small
$d_{0\ell}=\frac{\partial \phi}{\partial k_\ell} \approx 0$, which further simplifies
Eq. (\ref{eq:step})).

\subsection{Ramping Phase Modulation}
\label{sec:ramp}

In this Section we will discuss another type of phase modulation -- the case when the
phase is linearly changes with time. This modulation utilizes the  phase ramping
technique.(For more details, see Refs.\citeonline{colavita1}-\citeonline{shao}.) To
develop analytical solution we will be using the system equations developed above,
specifically Eqs.   (\ref{eq:dtintf}) and  (\ref{eq:Rli0lz}).

The optical path difference for the case of ramping phase modulation is modeled as a
continuous function of time as follows: 
\begin{equation}
x(t)=  x_0+ v\, t, 
\label{eq:time_p}
\end{equation}

\noindent where $x_0$ is the initial OPD value and $v$ is the
instantaneous velocity of OPD modulation. Remembering the definition for $\tau$ as
$\tau=t-t_i$,  and  $\Delta
\tau_i=t^+_i-t^-_i$ and $t_i=\frac{1}{2}(t^+_i+t^-_i)$, Eq. (\ref{eq:ugly04}) takes
the  form:
\begin{eqnarray}
{\tilde{\cal P}}_{\ell i}&=&
 {e^{ j\, k_\ell  x(t_i) }} \delta{\tilde{\cal P}}_{\ell i},
\label{eq:ugly03z} 
\end{eqnarray}
with coefficient $\delta{\tilde{\cal P}}_{\ell i} $ given by
\begin{eqnarray}
\delta{\tilde{\cal P}}_{\ell i}= \frac{1}{ \Delta\tau_i}
\int_{-\frac{1}{2}\Delta\tau_i}^{+\frac{1}{2}\Delta\tau_i} 
\!\! e^{ j\, k_\ell v\tau }
\,\frac{\sin[\frac{1}{2}\Delta k_\ell z(\tau)]}
{\frac{1}{2}\Delta k_\ell z(\tau)} \,d\tau  
\label{eq:ugly041} 
\end{eqnarray}
\noindent and $z(\tau)=d_{0\ell}+x(t_i)+v\,\tau$.
This allows us to present Eq. (\ref{eq:dtint9s}) in the following   form:
\small
\begin{eqnarray}
N_{\ell i} &=&   {\cal I}_{0 \ell} 
\bigg(1 +  V_{0\ell}\sin\big[\phi(k_\ell)+k_\ell x(t_i)\big] 
\,{\sf Re}\big\{\delta{\tilde{\cal P}}_{\ell i}\big\} + \nonumber\\
&&\hskip 31 pt +~
 V_{0\ell}\cos\big[\phi(k_\ell)+k_\ell x(t_i)\big] \, 
{\sf Im}\big\{\delta{\tilde{\cal P}}_{\ell i}\big\}\bigg)\!,
\hskip 10pt
\label{eq:dtint9zz}
\end{eqnarray}\normalsize

\noindent where  the complex matrix of additional rotation in the phase
space,  $\delta{\tilde{\cal P}}_{\ell i}$, is given by  
Eq. (\ref{eq:ugly04}). Equation (\ref{eq:dtint9zz})  may equivalently 
be presented in a matrix form as below 
\begin{eqnarray}
N_{\ell i}
&=& \Big(
1; ~~\sin k_\ell x(t_i);  ~~\cos k_\ell x(t_i) \Big) 
\times\nonumber\\[0pt]
&&\hskip 30pt \times 
\begin{array}({ccc}) 
1 &0 &  0\\[0pt]
0 & ~{\sf Re}\big\{\delta{\tilde{\cal P}}_{\ell i}\big\}& 
-{\sf Im}\big\{\delta{\tilde{\cal P}}_{\ell i}\big\}\\[0pt] 
0 &~{\sf Im}\big\{\delta{\tilde{\cal P}}_{\ell i}\big\}; &
~~{\sf Re}\big\{\delta{\tilde{\cal P}}_{\ell i}\big\} 
\end{array} 
\begin{array}({c})
{\cal I}_{0\ell}  \\[0pt]
{\cal I}_{0\ell} V_{0\ell}\cos  \phi(k_\ell) \\[0pt]
{\cal I}_{0\ell} V_{0\ell}\sin  \phi(k_\ell)  
\end{array}.
\label{eq:wj12b*}
\end{eqnarray}
\normalsize

The result of  integration of Eq.(\ref{eq:ugly041}) may not be presented in a compact
analytical form. It rather could be expressed in the form of two functions defined as
{\sf SinIntegral} and {\sf CosIntegral}. To simplify the analysis, the {\sf sinc}
function may be given in the form of power series expansion with respect to the small
parameter $\Delta k_\ell z(\tau)$ as given by Eq. (\ref{eq:sin_exp}). This expansion
allows us to  present   Eqs. (\ref{eq:ugly041}) in the following form: 
\begin{eqnarray}
\delta{\tilde{\cal P}}_{\ell i}&=&\!\!
\frac{1}{ \Delta\tau_i}
\int_{-\frac{1}{2}\Delta\tau_i}^{+\frac{1}{2}\Delta\tau_i} 
\!\! e^{ j\, k_\ell v\tau }
\,\frac{\sin[\frac{1}{2}\Delta k_\ell z(\tau)]}
{\frac{1}{2}\Delta k_\ell z(\tau)} \,d\tau =\nonumber\\[7pt]
&=&\!\!
\frac{1}{ \Delta\tau_i}
\int_{-\frac{1}{2}\Delta\tau_i}^{+\frac{1}{2}\Delta\tau_i} 
\!\! e^{ j\, k_\ell v\tau }
\,\Big(1-\frac{\Delta k_\ell^2 z(\tau)^2}{24}  +
\frac{\Delta k_\ell^4 z(\tau)^4}{1920} +
{\cal O}( \frac{\Delta k_\ell^6 z^6}{7!\,2^6})\Big)\, d\tau,
\label{eq:dP}
\end{eqnarray}
\noindent where $z(\tau)=d_{0\ell}+x(t_i) +v\,\tau= z_i+v\,\tau$ with
$z_i= d_{0\ell}+x(t_i)$.
This equation, (\ref{eq:dP}), was integrated to obtain the following result for 
$\delta{\tilde{\cal P}}_{\ell i}$:
\begin{eqnarray}
\delta{\tilde{\cal P}}_{\ell i}&=&\!\!
 \frac{\sin[\frac{1}{2} k_\ell\, v \,\Delta\tau_i]}
{\frac{1}{2} k_\ell\, v \,\Delta\tau_i} +
\tilde{\cal A}_{\ell i} \,\frac{\Delta k_\ell^2}{24 k_\ell^2}  + 
\tilde{\cal B}_{\ell i} \,
 \frac{\Delta k_\ell^4}{1920\,k_\ell^4} 
+ {\cal O}( \frac{\Delta k_\ell^6}{7!\,2^6k_\ell^6}), 
\hskip20pt
\label{eq:peli}
\end{eqnarray}

\noindent  where complex coefficients ${\cal A}_{\ell i}$ is
given as follows:
\begin{eqnarray}
\tilde{\cal A}_{\ell i}&=&\!\!
\Big[1+\big(1-jk_\ell z_i)\big)^2 
-(\frac{1}{2} k_\ell\, v \,\Delta\tau_i)^2\Big] \frac{\sin[\frac{1}{2}
k_\ell\, v \,\Delta\tau_i]} {\frac{1}{2} k_\ell\, v \,\Delta\tau_i}- \nonumber\\
&-&
2\Big(1-jk_\ell z_i\Big) \, 
 {\cos[\frac{1}{2} k_\ell\, v \,\Delta\tau_i]}  
\hskip35pt
\label{eq:Aeli}
\end{eqnarray}
and  ${\cal B}_{\ell i}$ is computed as follows:
\begin{eqnarray}
\tilde{\cal B}_{\ell i}&=&\!\!
\Bigg(\Big[1+\big(1-jk_\ell z_i\big)^2-
(\frac{1}{2} k_\ell\, v \,\Delta\tau_i)^2\Big]^2+ 
4\big(2-jk_\ell z_i\big)^2+
\nonumber\\[7pt]
&& \hskip 115pt+~
4\Big(1-(\frac{1}{2} k_\ell\, v \,\Delta\tau_i)^2\Big)\Bigg) \frac{\sin[\frac{1}{2}
k_\ell\, v \,\Delta\tau_i]} {\frac{1}{2} k_\ell\, v \,\Delta\tau_i}
-\nonumber\\[7pt] 
&-&
4\Bigg(5+(1-jk_\ell z_i)
\Big(\big(1-jk_\ell z_i\big)^2-
(\frac{1}{2} k_\ell\, v \,\Delta\tau_i)^2\Big)
\Bigg)
 {\cos[\frac{1}{2} k_\ell\, v \,\Delta\tau_i]}. 
\hskip20pt
\label{eq:Belia}
\end{eqnarray}
The obtained expressions may be used to simplify the results of
temporal integration Eq. (\ref{eq:dtint9zz}). As a result, the coefficients 
${\sf Re}\big\{\delta{\tilde{\cal P}}_{\ell i} \big\}$ and 
${\sf Im}\big\{\delta{\tilde{\cal P}}_{\ell i} \big\}$  in the fringe
equation Eq. (\ref{eq:dtint9zz}) may be written in the following form:
\begin{eqnarray}
{\sf Re}\big\{\delta{\tilde{\cal P}}_{\ell i} \big\}&=&\!\!
 \frac{\sin[\frac{1}{2} k_\ell\, v \,\Delta\tau_i]}
{\frac{1}{2} k_\ell\, v \,\Delta\tau_i}\Big[1 +
\Big(2-k^2_\ell z_i^2 -(\frac{1}{2} k_\ell\, v \,\Delta\tau_i)^2\Big)
 \,\frac{\Delta k_\ell^2}{24 k_\ell^2}\Big]-\nonumber\\[7pt]
&& \hskip 75pt-~
2\,{\cos[\frac{1}{2} k_\ell\, v \,\Delta\tau_i]}  
 \,\frac{\Delta k_\ell^2}{24 k_\ell^2}  + 
{\cal O}( 
 \frac{\Delta k_\ell^4}{5!2^4\,k_\ell^4}),
\hskip20pt
\label{eq:rep}
\end{eqnarray}
\begin{eqnarray}
{\sf Im}\big\{\delta{\tilde{\cal P}}_{\ell i} \big\}&=&\!\!
2k_\ell z_i \Big(-\frac{\sin[\frac{1}{2} k_\ell\, v \,\Delta\tau_i]}
{\frac{1}{2} k_\ell\, v \,\Delta\tau_i}+
 {\cos[\frac{1}{2} k_\ell\, v \,\Delta\tau_i]}\Big) 
 \,\frac{\Delta k_\ell^2}{24 k_\ell^2}  + 
{\cal O}( 
 \frac{\Delta k_\ell^4}{5!2^4\,k_\ell^4}),
\hskip20pt
\label{eq:imp}
\end{eqnarray}
\noindent with  $z_i=d_{0\ell}+x(t_i)\equiv d_{0\ell}+x_0+v\,t_i$.

The obtained expression models the photon flux detected at the CCD
for the case of rectangular bandpass filter and ramping phase modulation. It extends
the results obtained for the monochromatic case on the finite size of spectral
bandwidth.  This fact is indicated by the explicit dependency of the obtained result
on the width of a spectral channel $\Delta k_\ell$.

\section{Discussion and Future Plans}
\label{sec:summary}

The main objective of this paper has been to introduce the reader to the
concepts and the instrumental logic of the SIM astrometric observations,
especially as they relate to  estimation of the white
light fringe parameters. The set of formulae described herein will serve as the
kernel for the future mission analysis and simulations.
We have also developed a set of expressions that may be used for fringe
visibility and phase extraction  for both SIM science and guide interferometers. 
The obtained expressions depend on the  effective operational
wavelength of OPD modulation, the width of a particular  spectral channel $\Delta
k_\ell$ with the mean wavenumber $k_\ell$ and corresponding wavelength 
$\lambda_\ell$.   Our model accounts for a number of  instrumental and physical
effects and is able to compensate for a number of operational regimes. 

The logic of our method is straightforward: one first assumes the desirable
properties of the bandpass filter, then finds the corresponding envelope function, and
then applies the obtained  expressions  (which are valid for a generic case).  The
obtained solutions for the envelope function $W$ and, most specifically,
$\delta{\tilde{\cal P}}_{\ell i} $ may be directly substituted either in the
expression for the complex visibility phasors 
Eqs. (\ref{eq:phasorse2_g})-(\ref{eq:phasorse1_g_p}), or  into   equations for the 
visibility, amplitude and phase of the fringe, given by Eqs.(\ref{eq:phasorse3_g_p}). 
We applied this formalism to the case of  a rectangular bandpass (the obtained results
are given by  Eqs. (\ref{eq:a1})-(\ref{eq:phasors0_diam})). While the complex
visibility phasors are linear with respect to photon counts, the explicit expressions
for the fringe parameters are non-linear. This fact may be used to design specific
properties of unbiased fringe estimators for processing the white light data. 

Having developed algorithm to obtain the best estimates for the fringe parameters, Eqs.
(\ref{eq:phasorse3_g_p}) it is naturally to write down the expression that would allow
to estimate the group delay for the polychromatic case. It turns out that the
following expression $d=\frac{1}{ N}\sum^L_\ell ({\bar{\phi}^\diamond_\ell}/{ k_\ell})$
is sufficient to estimate the group delay for the SIM bandwidth to the required
accuracy of a few tens of picometers.\cite{mct} This single channel error expression
can be used to determine the group delay error when combining multiple spectral
channels of data via phase delay or group delay methods.  These errors will vary based
on the underlying assumptions, but a reasonable figure of merit to keep in mind is
that a random 1$\%$ error in the wavelength combined with the nominal 10 nm rms delay
requirement leads to approximately 60 pm of delay error when using 4 spectral channels
of data processed with a least squares algorithm for each channel.  This result scales
linearly with the wavelength error and the delay offset.  In our further work we will
numerically address the problem of unbiased estimators for the fringe phase,
visibility and group delay. This effort is currently underway.

Our analysis shows\cite{Turyshev2000,mct} that, while the model of the rectangular
bandpass filter is working quite well, for the `real life' one must account for the
effect of leakage of light. This effect due to the leakage of light onto the
studied spectrometer pixel of the detector from the adjacent pixels with different
wavenumbers. At this moment, it seams more appropriate that a combination of the
rectangular bandpass filter with additional effect of light leakage from the
adjacent pixels that must be included into the model of a CCD detector. The
corresponding analysis,  simulation results and implications for the instrument design
will be reported elsewhere. 
 
\appendix
\section{Two Definitions for the Fringe Phase}
\label{sec:app_phase}

In this Appendix we will address the issue of the mean phase definition which 
requires  some additional work. It is
tempting to define the mean phase as 
\begin{equation}
\phi_\ell=\frac{1}{\Delta k_\ell}
\int_{k^{-}_\ell}^{k^{+}_\ell} \!\!\hat{\cal I}_{0\ell}(k)\phi(k)dk 
~\equiv~
\frac{1}{{\cal I}_{0 \ell}\Delta k_{\ell}}
\int_{k^{-}_\ell}^{k^{+}_\ell} \!\!
{\cal F} (k){\cal I}_0(k)\phi(k)dk.
\label{eq:fj}
\end{equation}
However, one needs to relate this expression   to the
quantity $\phi(k_\ell)$, which is the phase value at a particular wavenumber. 
Assuming that phase $\phi(k)$ is a slow varying function of $k$ and,
as such, it may be expanded in a Taylor series as follows:  
 \begin{equation}
\phi(k)=\phi(k_\ell)+\frac{\partial \phi}{\partial k}\Big|_{k_\ell}
\hskip -5pt \, (k-k_\ell)+
\frac{1}{2}\frac{\partial^2 \phi}{\partial k^2}\Big|_{k_\ell}
\hskip -5pt \, (k-k_\ell)^2+{\cal O}(\Delta k_\ell^3).
\label{eq:phik}
\end{equation}

We can now substitute this formula directly in Eq.(\ref{eq:fj}), which results in 
\begin{eqnarray}
\phi_\ell&=&\phi(k_\ell)+ 
\frac{1}{2}\frac{\partial^2 \phi}{\partial k^2}\Big|_{k_\ell}
\!\!\,\!\frac{1}{{\cal I}_{0 \ell}\Delta k_{\ell}}
\int_{k^{-}_\ell}^{k^{+}_\ell} \!\!
{\cal F} (k){\cal I}_0(k)(k-k_\ell)^2dk+{\cal O}(\Delta k_\ell^3).
\label{eq:fl}
\end{eqnarray}
Or in other words, the phase value $\phi(k_\ell)$ at a particular
wavenumber $k_\ell$ is related to the mean phase $\phi_\ell$
within the spectral channel with width $\Delta k_\ell$ (i.e.  
Eq.(\ref{eq:fj}))  by the
following expression
\begin{eqnarray}
\phi(k_\ell)&=&\phi_\ell- \frac{1}{2}
\frac{\partial^2 \phi}{\partial k^2}\Big|_{k_\ell}
\!\Delta k_{\ell}^2 ~\mu^{(2)}_\ell +{\cal O}(\Delta k_\ell^3),
\label{eq:kj21}
\end{eqnarray}

\noindent where $\mu^{(2)}_\ell$ is the dimension-less second-order moment of
wavenumber distribution within the spectral channel of interest:
\begin{equation}
\mu^{(2)}_\ell=\frac{1}{{\cal I}_{0 \ell}\Delta k_{\ell}^3}
\int_{k^{-}_\ell}^{k^{+}_\ell} \!\!
{\cal F} (k){\cal I}_0(k)(k-k_\ell)^2\,dk.
\end{equation}

In the general case, when the higher order moments are considered, this expression
takes the following form:
\begin{eqnarray}
\phi(k_\ell)&=&\phi_\ell- \sum_{p=2}^{P}\frac{1}{p!}
\frac{\partial^p \phi}{\partial k^p}\Big|_{k_\ell}
\!\Delta k_{\ell}^p ~\mu^{(p)}_\ell +{\cal O}(\Delta k_\ell^p),
\label{eq:kj2}
\end{eqnarray}
\noindent with moments $\mu^{(p)}_\ell$ given as follows:
\begin{eqnarray}
\mu^{(0)}_\ell &=&1, \qquad \mu^{(1)}_\ell=0, \qquad
\mu^{(2)}_\ell=\frac{1}{{\cal I}_{0 \ell}\Delta k_{\ell}^3}
\int_{k^{-}_\ell}^{k^{+}_\ell} \!\!
{\cal F} (k){\cal I}_0(k)(k-k_\ell)^2dk,\\
\mu^{(p)}_\ell&=&\frac{1}{{\cal I}_{0
\ell}\Delta k_{\ell}^{p+1}}
\int_{k^{-}_\ell}^{k^{+}_\ell} \!\!
{\cal F} (k){\cal I}_0(k)(k-k_\ell)^pdk, \qquad 0<|\mu^{(p)}_\ell|<1,
\qquad \forall p.
\end{eqnarray}

Note that the wavenumbers within a spectral channel may be considered uniformly
distributed, thus    prompting to use $\phi(k_\ell)=\phi_\ell+{\cal O}(\Delta
k_\ell^2)$.  However, the knowledge of the second moment  $\mu^{(2)}_\ell$ 
may be important in combining the fringe solution for the whole operational
bandwidth. This question will be addressed elsewhere.

\section{
Approximation for the Complex Fringe Envelope Function}
\label{sec:appa}

Expression  for the fringe envelope function Eq. (\ref{eq:wj}), 
contains terms that are of the first and higher orders of phase variation
within the ${\ell}$-th spectral channel. We shell separate these terms   by
expanding   phase   $\phi(k)$ in the Taylor series around the mean
wavenumber $k_\ell$ as given by Eq. (\ref{eq:phik}).  
This  transforms the   argument in  Eq. (\ref{eq:wj}) as:
\begin{eqnarray}
\big[(k-k_\ell)x(t)+\phi(k)-\phi(k_\ell)\big] 
&=&
 (k-k_\ell) \big[\,x(t)+d_{0\ell}\big]+{\cal O}(\Delta k_\ell^2),
\label{eq:wj3}
\end{eqnarray}
\noindent  
where  $d_{0\ell}=\frac{\partial \phi}{\partial k}\Big|_{k_\ell}$ is the 
group delay  within the ${\ell}$-th channel. 
In the regime of small phase 
variations within the spectral channel 
$\Delta k_\ell \frac{\partial \phi}{\partial k}\Big|_{k_\ell}
\equiv \Delta k_\ell \,d_{0\ell}\ll 1,$ 
we  can expand the exponential argument   in the
expression  Eq. (\ref{eq:wj}) as given below
\begin{eqnarray}
&&\hskip -20pt {\sf exp} \Big\{j\,\Big[ (k-k_\ell) 
\Big(x(t)+d_{0\ell}\Big)+ {\cal O}(\Delta
k_\ell^2)\Big]\Big\}=\nonumber\\
&&\hskip 60pt =~\bigg\{1+j\, (k-k_\ell) \,d_{0\ell}+ 
{\cal O}(\Delta k_\ell^2)\bigg\}\, {\sf exp}
\Big\{j\, (k-k_\ell) x(t)\Big\}.
\label{eq:wj62}
\end{eqnarray}

This last expression may be used to re-write the phase-dependent envelope
function from Eq.(\ref{eq:wj})  as 
\begin{eqnarray}
{\tilde W_\ell}\big[\Delta k_\ell, \phi_\ell, x(t)\big]=\bigg\{1+ 
d_{0\ell}\frac{\partial  }{\partial x(t)} + 
{\cal O}(\Delta k_\ell^2)\bigg\} 
\int_{-\infty}^{+\infty}\hat{V}_{0\ell}(k)\,
e^{j\, (k-k_\ell)x(t)}\,dk.
\label{eq:wj70}
\end{eqnarray}
\normalsize

\noindent Defining the unperturbed  fringe envelope function (i.e. that is
un-affected by the phase variations  inside the spectral channel) as below
\begin{eqnarray}
{\tilde W_\ell}\big[\Delta k_\ell, x(t)\big]=
\int_{-\infty}^{+\infty}\hat{V}_{0\ell}(k)\,
e^{j\, (k-k_\ell)x(t)}\,dk.
\label{eq:wj8}
\end{eqnarray}

\noindent We may present  expression (\ref{eq:wj70}) for envelope
function in the following form:
\begin{eqnarray}
{\tilde W_\ell}\big[\Delta k_\ell, \phi_\ell, x(t)\big]&=& 
{\tilde W_\ell}\big[\Delta k_\ell, x(t)\big] +
d_{0\ell} \, W'_\ell \big[\Delta k_\ell, x(t)\big]+
{\cal O}(\Delta k_\ell^2),
\label{eq:wj10a}
\end{eqnarray}
\noindent where   superscript $'$ denotes  partial 
derivative with respect to OPD ${\delta }/{\delta x(t)}$.
 
 At this point we have established the functional dependency of the envelope
function, but for the immediate purposes we will be using a generic form for
this function, ${\tilde W_\ell}\big[\Delta k_\ell,   x(t)\big]$  presenting it
only by it's amplitude and phase:
\begin{eqnarray}
{\tilde W_\ell}\big[\Delta k_\ell,   x(t)\big]&=& {\cal E}_{\ell}\big[\Delta k_\ell,
x(t)\big] {\sf exp}\Big\{j\Omega_{\ell}\big[\Delta k_\ell,
x(t)\big]\Big\}\equiv {\cal E}_{\ell} \,
{\sf exp}\Big\{j\Omega_{\ell}\Big\}.
\label{eq:wj11a}
\end{eqnarray}

\noindent Similarly to the expression (\ref{eq:wj10a}) we re-write
this result in the following form
\begin{eqnarray}
{\tilde W_\ell}\big[\Delta k_\ell, \phi_\ell, x(t)\big]&=& 
{\cal E}_{\ell}\, {\sf exp}\Big\{j\Omega_{\ell}\Big\} + 
d_{0\ell}\,\Big({\cal E}'_\ell + j {\cal E}_{\ell}\Omega'_\ell\Big) {\sf
exp}\Big\{j\Omega_{\ell}\Big\} +  {\cal O}(\Delta k_\ell^2). \hskip 10pt
\label{eq:ef}
\end{eqnarray}
At this moment we show the functional form of real and imaginary components of the
envelope function. Thus, from Eq. (\ref{eq:ef}) one immediately has
\begin{eqnarray}
{\sf Re}\Big\{{\tilde W}_\ell\big[\Delta k_\ell, \phi_\ell,x_i\big]\Big\} &=&
 {\cal E}_\ell \cos\Omega_{\ell}+d_{0\ell}\Big({\cal E}_\ell'
\cos\Omega_{\ell}- {\cal E}_\ell\Omega'_{\ell}\, 
\sin\Omega _{\ell}\Big)+{\cal O}(\Delta k^2_\ell), \hskip 10pt
\label{eq:wj12a*}\\[10pt]
{\sf Im}\Big\{{\tilde W}_\ell\big[\Delta k_\ell,\phi_\ell,x_i\big]\Big\}
&=& {\cal E}_\ell\sin \Omega_{\ell}+
d_{0\ell}\Big({\cal E}_\ell' \sin \Omega_{\ell}+
 {\cal E}_\ell\Omega'_{\ell}\, 
\cos \Omega_{\ell}\Big) +{\cal O}(\Delta k^2_\ell).\hskip 10pt
\label{eq:wj12b*+}
\end{eqnarray}

\noindent The obtained equation exhibits  explicit dependence on the phase variation
inside the spectral channel given by  $d_{0\ell}= {\partial \phi}/{\partial
k}\big|_{k_\ell}$. This issue will be addressed elsewhere.

\section{Solution for Rectangular Bandpass  
and Stepping OPD Modulation}
\label{sec:ow_rec}

In consideration of completeness, we present here a general
case solution for an optimally-weighted visibility phasor  for a rectangular
bandpass and stepping OPD modulation.  In the previous Section we obtained  this
solution in a general case, therefore,  the desired solution may be obtained directly
with the help of expressions  (\ref{eq:phasorse1_g_p}). Corresponding
optimally-weighted  solution may be presented  in the  form of Eqs.
(\ref{eq:phasorse1_g_p}) and (\ref{eq:phasorse3_g_p}) with coefficients ${\cal
A}^\diamond_{  k},  {\cal B}^\diamond_{k},  {\cal C}^\diamond_{ k}$  and ${\cal
D}^\diamond$  depend only on  the size of modulation steps $x_i$, mean  wavenumber
$\bar k$,   width of a spectral channel  $\Delta k$, variances of the data
$\sigma^2_i$  in a particular temporal bin. These coefficients are  given as
follows:
\begin{eqnarray} 
{\cal A}^\diamond_k &=&
\overset{N}{\underset{ij}\sum}~\frac{1}{\sigma_i^2\sigma_j^2\sigma_k^2} \,
{\rm sinc}[\frac{\Delta k x_i}{2}] \,
{\rm sinc}[\frac{\Delta k x_j}{2}]\,
\sin[\bar{k} (x_i- x_j)]\times
\nonumber\\[2pt]
&&\hskip 33pt \times~\Big[{\rm sinc}[\frac{\Delta k x_i}{2}]\,
{\rm sinc}[\frac{\Delta k x_j}{2}]\,
\sin[\bar{k} (x_i- x_j)]+\nonumber\\[2pt]
&&\hskip 40pt +~
{\rm sinc}[\frac{\Delta k x_j}{2}]\,
{\rm sinc}[\frac{\Delta k x_k}{2}]\sin[\bar{k}  (x_j- x_k)]+
\nonumber\\[2pt]
&&\hskip 40pt +~
{\rm sinc}[\frac{\Delta k x_k}{2}]\,
{\rm sinc}[\frac{\Delta k x_i}{2}]\,\sin[\bar{k} (x_k- x_i)]\Big],
\label{eq:a1}
\end{eqnarray} 
\begin{eqnarray} 
{\cal B}^\diamond_k&=&
\overset{N}{\underset{ij}\sum}~\frac{1}{\sigma_i^2\sigma_j^2\sigma_k^2}  \Big( 
{\rm sinc}[\frac{\Delta k x_i}{2}]\cos \bar{k}x_i-
{\rm sinc}[\frac{\Delta k x_j}{2}]\cos \bar{k}x_j\Big)\times
\nonumber\\[2pt]
&&\hskip 33pt \times~
\Big[{\rm sinc}[\frac{\Delta k x_i}{2}]\,
{\rm sinc}[\frac{\Delta k x_j}{2}]\,
\sin[\bar{k} (x_i- x_j)]+\nonumber\\[2pt]
&&\hskip 40pt +~
{\rm sinc}[\frac{\Delta k x_j}{2}]\,
{\rm sinc}[\frac{\Delta k x_k}{2}]\,\sin[\bar{k}  (x_j- x_k)]+
\nonumber\\[2pt]
&&\hskip 40pt +~
{\rm sinc}[\frac{\Delta k x_k}{2}]\,
{\rm sinc}[\frac{\Delta k x_i}{2}]\,\sin[\bar{k} (x_k- x_i)]\Big],
\end{eqnarray} 
\begin{eqnarray} 
{\cal C}^\diamond_k&=&-~
\overset{N}{\underset{ij}\sum}~\frac{1}{\sigma_i^2\sigma_j^2\sigma_k^2}  \big(
{\rm sinc}[\frac{\Delta k x_i}{2}]\,\sin \bar{k}x_i-
{\rm sinc}[\frac{\Delta k x_j}{2}]\,\sin \bar{k}x_j\big)\times
\nonumber\\[2pt]
&&\hskip 33pt \times~
\Big[{\rm sinc}[\frac{\Delta k x_i}{2}]\,
{\rm sinc}[\frac{\Delta k x_j}{2}]\, \sin[\bar{k} (x_i- x_j)]+\nonumber\\[2pt]
&&\hskip 40pt +~
{\rm sinc}[\frac{\Delta k x_j}{2}]\,
{\rm sinc}[\frac{\Delta k x_k}{2}]\,\sin[\bar{k}  (x_j- x_k)]+
\nonumber\\[2pt]
&&\hskip 40pt +~
{\rm sinc}[\frac{\Delta k x_k}{2}]
{\rm sinc}[\frac{\Delta k x_i}{2}]\sin[\bar{k} (x_k- x_i)]\Big],
\end{eqnarray} 
\begin{eqnarray} 
{\cal D}^\diamond &=&
\overset{N}{\underset{k}\sum}~{\cal A}^\diamond_{ k}~=\nonumber\\[2pt]
&=&\frac{1}{3} 
\overset{N}{\underset{ijk}\sum}~\frac{1}{\sigma_i^2\sigma_j^2\sigma_k^2}
\Big[{\rm sinc}[\frac{\Delta k x_i}{2}] {\rm sinc}[\frac{\Delta k x_j}{2}]
\sin[\bar{k} (x_i- x_j)]+\nonumber\\[2pt]
&&\hskip 33pt +~
{\rm sinc}[\frac{\Delta k x_j}{2}]\,
{\rm sinc}[\frac{\Delta k x_k}{2}]\,\sin[\bar{k}  (x_j- x_k)]+
\nonumber\\[2pt]
&&\hskip 33pt +~
{\rm sinc}[\frac{\Delta k x_k}{2}]\,
{\rm sinc}[\frac{\Delta k x_i}{2}]\,\sin[\bar{k} (x_k- x_i)]\Big]^2,
\hskip 10pt
\label{eq:phasors0_diam}
\end{eqnarray}
\noindent where, in consideration of brevity, we omitted  index $\ell$ denoting a
particular spectral channel. 

The obtained result Eqs. (\ref{eq:phasors0_diam})
clearly depends on the {\sf sinc} envelope function   and thus it has all the
information that is necessary  for the phase estimation purposes in the case of
the wide band-pass.  Note, that this result assumes that the phase does not
change inside the spectral channel.  Also, this result directly corresponds to
the result obtained for the monochromatic case. This may be demonstrated by
taking the limit  $\Delta k/{\bar k} \rightarrow 0$, which will lead to recovering
the familiar form of monochromatic fringe with coefficients ${\cal A}^\diamond_{k},
{\cal B}^\diamond_k,  {\cal C}^\diamond_k$  and ${\cal D}^\diamond$    given as
follows: 
\begin{eqnarray} 
{\cal A}_{k}^\diamond &=&
 \overset{N}{\underset{ij}\sum}~\frac{1}{\sigma_i^2\sigma_j^2\sigma_k^2}
\sin[k(x_i-x_j)] \times\nonumber\\
&&\hskip 50pt \times
\Big[\sin[k (x_i- x_j)]+\sin[k (x_j- x_k)]+
\sin[k (x_k- x_i)]\big],
\label{eq:phasorse2_ga1}
\end{eqnarray} 
\begin{eqnarray} 
{\cal B}_{k}^\diamond&=&
 \overset{N}{\underset{ij}\sum}~\frac{1}{\sigma_i^2\sigma_j^2\sigma_k^2}
\big( \cos kx_i-\cos kx_j\big)~ 
\times\nonumber\\
&&\hskip 50pt \times
\Big[\sin[k (x_i- x_j)]+\sin[k (x_j- x_k)]+\sin[k (x_k- x_i)]\Big],
\end{eqnarray} 
\begin{eqnarray} 
{\cal C}_{k}^\diamond &=& -~ 
 \overset{N}{\underset{ij}\sum}~\frac{1}{\sigma_i^2\sigma_j^2\sigma_k^2}
\big( \sin kx_i-\sin kx_j\big)~ \times\nonumber\\
&&\hskip 50pt \times
\Big[\sin[k (x_i- x_j)]+\sin[k (x_j- x_k)]+\sin[k (x_k- x_i)]\Big],
\end{eqnarray} 
\begin{eqnarray} 
{\cal D}^\diamond&=&\overset{N}{\underset{k}\sum}~{\cal A}^\diamond_{k}=
\nonumber\\
&=& 
\frac{1}{3}\overset{N}{\underset{ijk}\sum}\frac{1}{\sigma_i^2\sigma_j^2\sigma_k^2}
\big[\sin[k (x_i- x_j)]+\sin[k (x_j- x_k)]+\sin[k (x_k- x_i)]\big]^2.
\label{eq:phasorse2_ga}
\end{eqnarray}
 
This form  demonstrates that only the terms with $i\not=j\not=k\not=i$ are producing a
non-zero contributions to the result, while the terms where at least two of the
indexes are equal (i.e. $i=j$ or $i=k$ or $j=k$) will vanish from the sum.  
 
\subsection{The Optimally Weighted 4-bin (ABCD) Algorithm}
\label{sec:4bmon_g}

Results obtained in the previous  Section are valid for any number of OPD modulation
steps. In order to show their correspondence to  well-known formulations we
will present their particular form for the case with $N=4$,  equal
step sizes and wavelength-matched OPD modulation strokes. This will allow us to obtain
results that are constituting a four-bin algorithm. 

Thus, in  the case of 4 equal step sizes all of  $\frac{1}{4}\lambda$, the OPD is
stepping in increments of $x_i=\frac{\lambda}{4} i, ~  i\in[1,...,4]$ that correspond
to the phase changing  in steps of $\phi_i=\frac{\pi}{2} i$. By calculating the
quantities ${\cal A}^\diamond_{k},
{\cal B}^\diamond_k,  {\cal C}^\diamond_k$  and ${\cal D}^\diamond$  in Eqs.
(\ref{eq:phasorse2_ga1})-(\ref{eq:phasorse2_ga}) and substituting them into  Eq.
(\ref{eq:phasorse3_g_p}) we recover the optimally-weighted  form of  4-bin (or ABCD)
algorithm:
\vskip -10pt  
 \small
\begin{eqnarray} 
{\cal I}_0~V\cos\phi_0&=& 
\frac{ {\bar N}_1\Big(\sigma^2_2+2\sigma^2_3+\sigma^2_4\Big)-
{\bar N}_3\Big(2\sigma^2_1+\sigma^2_2+\sigma^2_4\Big)+
\big({\bar N}_2+{\bar N}_4\big)\Big(\sigma^2_1-\sigma^2_3\Big)}
{2\Big(\sigma^2_1+\sigma^2_2+\sigma^2_3+\sigma^2_4\Big)}, 
\nonumber\\[12pt]
{\cal I}_0~V\sin\phi_0&=&  
\frac{ {\bar N}_4\Big(\sigma^2_1+2\sigma^2_2+\sigma^2_3\Big)-
{\bar N}_2\Big(\sigma^2_1+\sigma^2_3+2\sigma^2_4\Big) +
\big({\bar N}_1+{\bar N}_3\big)\Big(\sigma^2_4-\sigma^2_2\Big)}
{2\Big(\sigma^2_1+\sigma^2_2+\sigma^2_3+\sigma^2_4\Big)},
\nonumber\\[12pt]
{\cal I}_0 &=&  
\frac{ \big({\bar N}_1+{\bar N}_3\big)\Big(\sigma^2_2+\sigma^2_4\Big)+
\big({\bar N}_2+{\bar N}_4\big)\Big(\sigma^2_1+\sigma^2_3\Big) }
{2\Big(\sigma^2_1+\sigma^2_2+\sigma^2_3+\sigma^2_4\Big)}.
\end{eqnarray} 
\normalsize

\noindent This solution produces following results for the 
optimally-weighted visibility, $V^2_\diamond$,
and the  phase, $\phi_0^\diamond$:

\small
\begin{eqnarray} 
V^2_\diamond &=&    \frac{
\begin{array}{l}
\Big[{\bar N}_1\Big(\sigma^2_2+2\sigma^2_3+\sigma^2_4\Big)-
{\bar N}_3\Big(2\sigma^2_1+\sigma^2_2+\sigma^2_4\Big)+
\big({\bar N}_2+{\bar N}_4\big)\Big(\sigma^2_1-\sigma^2_3\Big)\Big]^2+
\nonumber\\[10pt]
\hskip 10pt +~\Big[{\bar N}_4\Big(\sigma^2_1+2\sigma^2_2+\sigma^2_3\Big)-
{\bar N}_2\Big(\sigma^2_1+\sigma^2_3+2\sigma^2_4\Big) +
\big({\bar N}_1+{\bar N}_3\big)\Big(\sigma^2_4-\sigma^2_2\Big)\Big]^2
\nonumber\\[5pt]
\end{array}}
{\Big[\big({\bar N}_1+{\bar N}_3\big)\Big(\sigma^2_2+\sigma^2_4\Big)+
\big({\bar N}_2+{\bar N}_4\big)\Big(\sigma^2_1+\sigma^2_3\Big)\Big]^2},
\nonumber\\[12pt]
\phi_0^\diamond &=&  {\sf ArcTan} \left[
\frac{ {\bar N}_4\Big(\sigma^2_1+2\sigma^2_2+\sigma^2_3\Big)-
{\bar N}_2\Big(\sigma^2_1+\sigma^2_3+2\sigma^2_4\Big) +
\big({\bar N}_1+{\bar N}_3\big)\Big(\sigma^2_4-\sigma^2_2\Big)}
{{\bar N}_1\Big(\sigma^2_2+2\sigma^2_3+\sigma^2_4\Big)-
{\bar N}_3\Big(2\sigma^2_1+\sigma^2_2+\sigma^2_4\Big)+
\big({\bar N}_2+{\bar N}_4\big)\Big(\sigma^2_1-\sigma^2_3\Big)}\right].
\hskip 20pt
\label{eq:4bin_mg}
\end{eqnarray}
\normalsize

This is the optimally weighed form of the most well-known fringe estimation ABCD
algorithm. The absence of extra $\frac{\pi}{4}$ phase in the result
$\phi_0$ is  due to the phase-stepping approach.\cite{creath,greivenkamp}  In the case
when experimental noise is absent, this result directly correspond to the previously
obtained solutions for the ABCD algorithm in the monochromatic case.    

\section*{Acknowledgement}
The author acknowledges many useful discussions with Mark Colavita, Mike Shao, and
Mark Milman on several topics in this paper, especially their suggestion for
averaging phasors to eliminate bias. The reported research   has been done at the
Jet Propulsion Laboratory,  California Institute of Technology, which is under 
contract to the  National Aeronautic and Space Administration.

\end{document}